\documentclass[10pt,aps, twocolumn]{revtex4-2}

\usepackage{hyperref}
\usepackage{amsmath,amssymb}
\usepackage{graphicx}
\usepackage{xcolor}
\usepackage{siunitx}
\newcommand{\nocontentsline}[3]{}
\newcommand{\tocless}[2]{\bgroup\let\addcontentsline=\nocontentsline#1{#2}\egroup}

\newcommand{\figref}[1]{Figure~\ref{#1}}
\newcommand{\supfigref}[1]{Appendix Figure~\ref{#1}}

\newcommand{\fs}[1]{#1}
\newcommand{\ccbin}{\ensuremath{CC_{\text{bin}}}}
\newcommand{\ccwei}{\ensuremath{CC_{\text{w}}}}

\graphicspath{{figures/}}

\begin{document}
\title{Revising clustering and small-worldness in brain networks}
\author{Tanguy Fardet$^{1,2,3}$}
\author{Emmanouil Giannakakis$^{1, 2}$}
\author{Lukas Paulun$^{4,5}$}
\author{Anna Levina$^{1,2}$}
\affiliation{
  $^1$ University of T\"ubingen, Maria-von-Linden Stra\ss e 6, 72076 T\"ubingen, Germany,\\
  $^2$ \mbox{Max Planck Institute for Biological Cybernetics, Max Planck Ring 8, 72076 T\"ubingen, Germany}, \\
  $^3$ LEESU, École des Ponts, Université Paris Est Créteil, Marne-la-Vallée, France, \\
  $^4$ Bernstein Center Freiburg, Hansastra\ss e 9a, 79106 Freiburg, Germany, \\
  $^5$ University of Freiburg, Friedrichstra\ss e 39, 79098 Freiburg, Germany
}


\begin{abstract}
As more connectome data become available, the question of how to best analyse the structure of biological neural networks becomes increasingly pertinent. In brain networks, knowing that two areas are connected is often not sufficient, as the directionality and weight of the connection affect the dynamics in crucial ways. 
Still, the methods commonly used to estimate network properties, such as clustering and small-worldness, usually disregard features encoded in the directionality and strength of network connections. 
To address this issue, we propose using fully-weighted and directed clustering measures that provide higher sensitivity to non-random structural features. Using artificial networks, we demonstrate the problems with methods routinely used in the field and how fully-weighted and directed methods can alleviate them. Specifically, we highlight their robustness to noise and their ability to address thresholding issues, particularly in inferred networks.
We further apply our method to the connectomes of different species and uncover regularities and correlations between neuronal structures and functions that cannot be detected with traditional clustering metrics. 
Finally, we extend the notion of small-worldness in brain networks to account for weights and directionality and show that some connectomes can no longer be considered ``small-world''.
Overall, our study makes a case for a combined use of fully-weighted and directed measures to deal with the variability of brain networks and suggests the presence of complex patterns in neural connectivity that can only be revealed using such methods. 
\end{abstract}

\maketitle

\setcounter{secnumdepth}{3}
\setcounter{tocdepth}{3}
\tableofcontents

\newpage

\section{Introduction}

\fs{Comprehension of the brain connectome at different scales is essential for understanding neural systems.}
The structure of brain networks can provide valuable insights into their dynamical behaviours and functions, reflecting the ubiquitous relation between form (structure) and function (dynamics) \cite{Wang.2016}.
Today, there is an abundance of structural and functional \cite{Bonifazi.2019, Brookes.2011} connectivity data at different scales derived by various methods, ranging from microscale connectomes, which map the connections between individual neurons \cite{Cook.2019, Scheffer.2020, Winding.2023}, mesoscale connectomes that capture local circuits like cortical columns \cite{Oh.2014}, and macroscale networks, where nodes correspond to entire brain areas \cite{Craddock.2013}. 

\fs{To understand this massive amount of data, network science has developed a variety of measures to extract essential features of different network structures and find similarities between them.}
The abstract mathematical tools of network science are blind to the particular meanings of nodes and connections in the real world.
Therefore, they allow the use of the same analysis techniques to a wide range of different networks and enable comparisons between them \cite{Bassett.2017}.

\fs{Clustering is a particularly important metric to characterize recurrent interactions between neurons or brain regions through redundancy and shared information transfer.}
In growing networks, we expect specific clustering properties related to preferential attachment and the segregation of inputs and outputs \cite{Newman.2001} to emerge.
Furthermore, clustering metrics are involved in the computation of higher-order graph properties, such as small-worldness that captures a graph's propensity to form short average path lengths while maintaining high clustering~\cite{Muldoon2016}.

\fs{Brain networks, at all scales, are intrinsically directed and weighted, with weights often following heavy-tailed distributions.}
Nonetheless, many network measures, including clustering, are often analysed using the undirected or binary version of the graphs in many standard toolboxes~\cite{Conn,GRETNA,GraphVar}, ignoring directionality, connection weights, or both.
A single toolbox~\cite{BrainConnectivityToolbox} includes directed weighted measures, and it only computes the sum of all directed clustering coefficients (called ``total'' clustering), which does not provide significant additional information compared to the undirected coefficient.
Because the binary clustering and the partially-weighted methods implemented in these toolboxes are very sensitive to spurious edges, computing the clustering coefficient requires pre-processing in the form of thresholding in which low-weight edges are removed, an arbitrary process that often leads to ambiguous results.
For binary methods, all edges in the remaining network are then treated as having the same weight. 

\fs{Removing spurious edges through thresholding is often necessary to recover an approximation of the ``ground truth'' graph, as most brain networks cannot be measured directly but are inferred from indirect observations.}
During the inference process, noise and statistical biases can lead to the emergence of spurious low-weight edges.
Although a wide range of methods for thresholding have been suggested \cite{Bielczyk.2018, Stam.2014, Dimitriadis.2017}, there is no agreement on the best method for any given network.
Furthermore, as the weight distributions of spurious and real edges often overlap, perfectly discriminating edges based on weights alone is impossible.
This is concerning since spurious edges can have large effects on the network analysis, especially if weights are ignored, and spurious low-weight edges are treated the same as real edges.

\fs{In this paper, we propose a solution to the spurious edges and thresholding problems using directional and weighted clustering methods.}
Instead of focusing on the difficult task of finding a sensible threshold for a given network, we suggest the use of weighted network measures that fulfil what we call the \emph{continuity condition}; 
This condition requires that a measure should remain as robust as possible if a network is perturbed by the addition or deletion of weak edges and that an edge with infinitesimally small weight should be equivalent to the absence of that edge. 
We discuss weighted directed clustering coefficients that fulfil the continuity condition and demonstrate how they remove the need for thresholding weighted connectomes altogether.
Furthermore, in the second part of the paper, we show that these methods correlate with biological and structural features in several connectomes and are useful to find and analyse new properties linking structure and function in specific circuits. We particularly identify structural features in the mouse connectome that cannot be identified via commonly used hybrid clustering methods.
Finally, we define a weighted directed measure using these continuous clustering methods to assess the small-world propensity of connectomes.
Using our metric, we show that accounting for both weights and directedness leads to results that challenge the notion that small-worldness is an intrinsic feature of nervous systems and brain networks.


\section{Results}


In the following sections, we argue that taking directionality and connection strength into account is a necessary step for reliable network analysis in neuroscience. We further compare different methods of estimating the weighted clustering coefficient and argue that fully weighted definitions provide better insights into network structure. 

In the first part of the results, we use artificial networks to demonstrate the importance of using weighted and directed measures and compare different methods. We then use connectomics data from four species and illustrate the strong effect that different clustering definitions have on the analysis of real neuroscientific data.

\subsection{Artificial Networks}

\subsubsection{Using undirected measures leads to information loss}

\begin{figure}
	\includegraphics[width=\columnwidth]{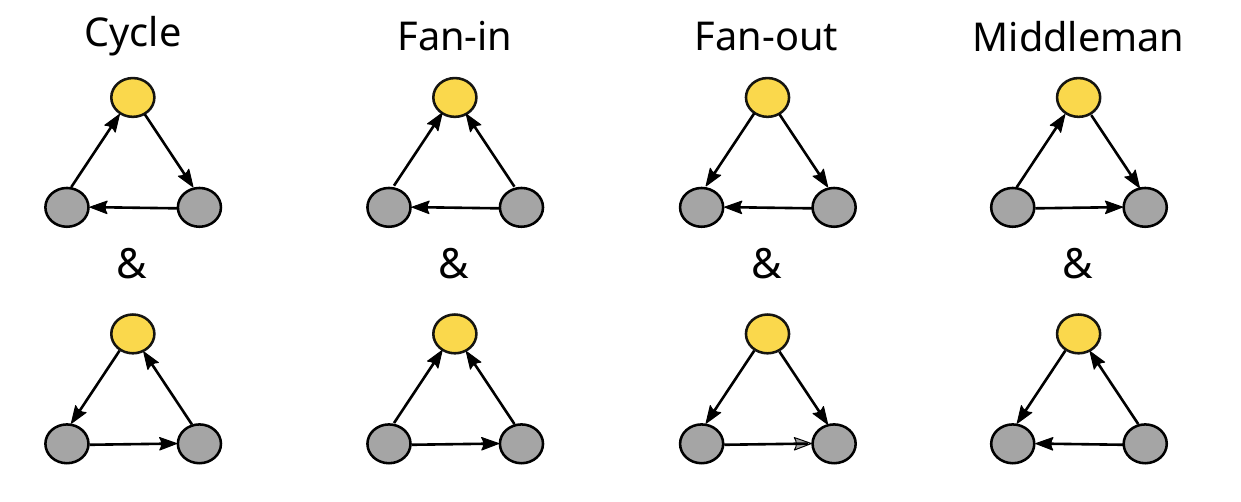}
	\caption{Illustration of the four directed clustering motifs. Each possible version of the cycle, fan-in, fan-out and middleman motifs are shown for the top node (in yellow).}
	\label{fig:motifs}
\end{figure}

\begin{figure*}
	\includegraphics[width=\textwidth]{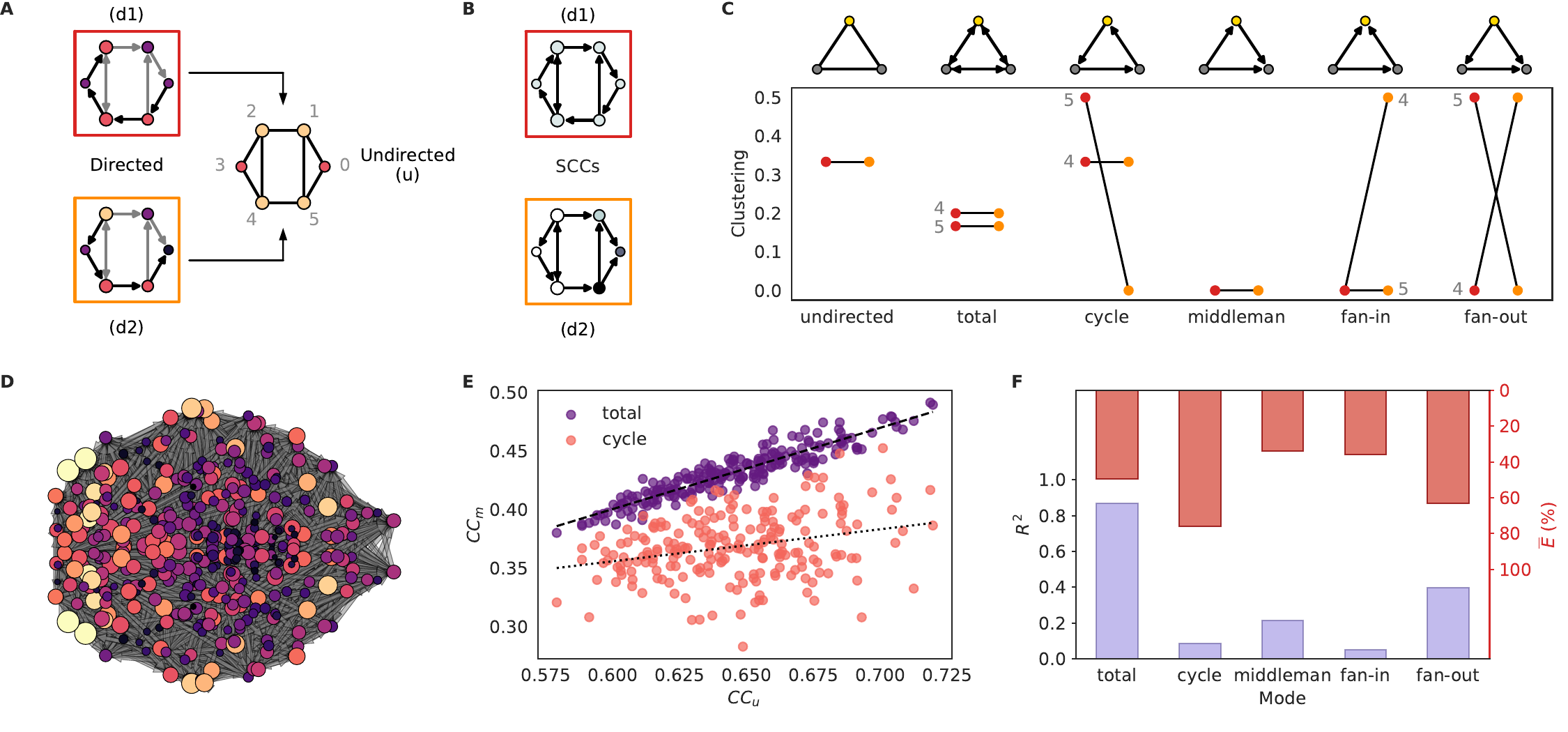}
	\caption{
		Using the undirected representation of directed graphs, or undirected measures lead to a significant loss of information.
		\textbf{A}. Two different directed networks correspond to the same undirected network (differing edges are highlighted in black).
		Node size gives the total degree, and colour gives the out-degree (low in black to high in yellow).
		Associated node IDs are shown in grey around the undirected version of the graphs.
		\textbf{B}. Comparison of the strongly connected components (SCCs, given by node colours) and motifs of the two directed graphs.
		(d1, dark red) is strongly connected with each node reachable from each other, whereas (d2, orange) falls apart into 4 distinct SCCs.
		This means that if information is injected in 5 for (d2), it can only reach nodes 0 and 1, whereas it can reach all other nodes in (d1).
		\textbf{C}. Comparison of the undirected and directed clustering coefficients for nodes 4 and 5 and their evolution (represented by the black line) across graphs d1 (dark red) and d2 (orange).
		The triangular motifs are represented above (for a chosen node, in yellow); the total clustering considers all possible variations of the directed triangles.
		Though they seem identical on the undirected graph, nodes 4 and 5 actually differ (either within a single graph or across graphs) for all directed clustering motifs except middleman.
		\textbf{D}. Mouse brain (same rules as A).
		\textbf{E}. Linear regression between the undirected clustering ($CC_u$) and two directed motifs ($CC_m$).
		\textbf{F}. $R^2$ and mean absolute percentage error $\overline{E}$ associated to the regression between the undirected clustering and each directed motif.}
	\label{fig:comparison_directed_undirected}
\end{figure*}

Directionality critically affects how information flows in the brain, and thus measures developed for directed connectomes should be used. However, many network metrics used in neuroscience disregard directionality \cite{Kale2018}, as attested by the fact that multiple neuroscience toolboxes implement only the undirected form of the clustering coefficient (e.g.  \texttt{Conn}, \texttt{GRETNA}, or \texttt{GraphVar}~\cite{Conn,GRETNA,GraphVar}, totalling more than 5000 citations).
It might seem that undirected clustering already captures all relevant information and is difficult to extend to directed measures, but in this section, we argue that both of these statements are not true.   

The classical unweighted clustering coefficient definition ($CC_u$) for undirected networks simply assesses how close the neighbours of a node are to being a fully connected graph. On a practical level, the computation of the $CC_u$ assesses the probability of connection with neighbouring nodes  by looking at all the triangles formed by the node and two of its neighbours normalized by the number of possible triangles. 
To extend the definition to directed networks, we can separately concentrate on different types of possible triangles. In general, there 8 different types of possible directed triange  (each link can be oriented two different ways; here, we count bidirectional links as two different links that contribute to their own set of triangles). However, for reasons of symmetry, we need to consider only four different motifs: cycle, middleman, fan-in, and fan-out --- \figref{fig:motifs}. To find the corresponding clustering coefficient for each pattern, we simply assess how often the particular type of triangle is present, normalized by the theoretically possible number of triangles given the links from a node to its neighbours (for exact definition, see Methods~\ref{methods-clustering}).

We first illustrate how the directional clustering coefficient can discern the differences between nodes in the network. For this, we look at simple generated networks that differ only in the directionality of four links,  \figref{fig:comparison_directed_undirected}A. The first observation is that the network (d1, red) is strongly connected, meaning that information injected in any node of the network can reach any other node. On the other hand, network (d2, orange) has 4 distinct strongly connected components. This means that if information is injected in node 5 of (d2), it can only reach nodes 0 and 1, whereas it can reach all other nodes in (d1) --- \figref{fig:comparison_directed_undirected}B. 

How much of the differences between these two networks can be picked up by the clustering coefficient? As expected, there are no differences when using undirected or total clustering.
However, we can obtain deeper insights about information transmission in the networks by considering specific motifs separately --- \figref{fig:comparison_directed_undirected}C. For example, directed motifs expose significant differences between nodes 4 and 5. 

To check whether there are similar differences in the information provided by specific directed clustering motifs in real neuronal connectomes, we studied the unweighted mesoscopic brain network of the mouse~\cite{Oh.2014} --- \figref{fig:comparison_directed_undirected}D. 
Looking only at the total clustering, it may seem that undirected clustering already captures most of the information, as they appear strongly correlated in \figref{fig:comparison_directed_undirected}E.
However, the picture is already quite different for the cycle motif and we show that only a small fraction of the variance (between $0.05$ for fan-in and $0.4$ for fan-out) can be captured when trying to predict the clustering values of any specific motif from the undirected local clustering --- \figref{fig:comparison_directed_undirected}F.
Thus, additional, complementary information in the different motif-specific clustering coefficients exists, which may lead to novel discoveries when analysing the properties of various nodes (be they neurons or brain areas) together with their function.


\subsubsection{Fully-weighted methods are more sensitive to weight contributions}

In biological neural networks, the connection strength carries a significant part of the information about the interactions between nodes. 
Multiple extensions of measures originally developed for binary networks have been proposed to adequately capture features of weighted networks. We recently suggested~\cite{fardetWeightedDirectedClustering2021} to classify these definitions based on their relationship with \emph{binary-based properties} (properties that can be computed directly from the adjacency matrix) in two broad categories. \emph{Hybrid methods} that use both weights and node degrees (a binary property) to compute the clustering coefficient, such as the methods of  Barrat~\textit{et al.}~\cite{Barrat.2004}, Onnela~\textit{et al.}~\cite{Onnela.2005} and their generalizations to directed networks \cite{Clemente.2018, Fagiolo.2007}
and \emph{fully-weighted methods} that rely only on the connection weights to compute the clustering coefficient, such as the method of Zhang and Horvath \cite{Zhang.2005} and the continuous method ~\cite{fardetWeightedDirectedClustering2021}.

\begin{figure*}
	\includegraphics[width=\textwidth]{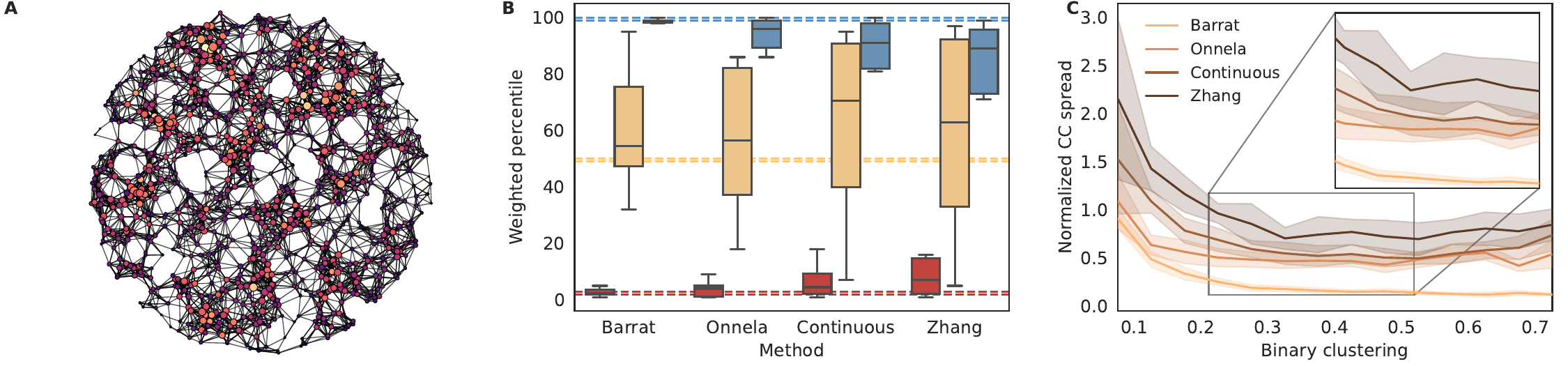}
	\caption{Fully-weighted methods are more sensitive to the weights-related features of the nodes. 
		\textbf{A}. 1000-node spatial network with distance-dependent connectivity and weights drawn from a log-normal distribution. 
		Node size indicates the total degree, node color indicates out-degree.
		\textbf{B}. Difference between binary and weighted middleman clustering ranks of the nodes from the previous graph.
		Red: 2nd percentile, yellow: 50th percentile, blue 100th percentile.
  The theoretical locations of the associated binary percentiles are marked by dashed lines of the corresponding colours).
		While Barrat and Onnela's methods remain relatively close to the binary ranking, notably for extreme values, the fully weighted methods significantly deviate from it even for the lowest and highest percentiles.
		\textbf{C}. 20 realizations of the spatial network in A, with weights sampled from the same lognormal distribution, are used to make groups of nodes with similar binary clustering coefficients (in bins of 0.05 width).
		Normalized $CC_w$ spread $s$ is computed as $s = IQ / M$, with $IQ$ the interquartile-range and $M$ the median.
		Shaded areas give the standard deviation around the mean and are obtained via bootstrapping.
		The ratio between low and high clustering values is higher for fully weighted clustering methods, meaning that their discriminating power is higher than that of the hybrid methods.
		The inset is a zoom on the values containing 90\% of the measured nodes.
		See Methods \ref{methods:analysis} for details about the network, weights, and analysis.
	}
	\label{fig:weight_contrib}
\end{figure*}

To demonstrate the increased sensitivity of fully-weighted methods, we consider a spatial network reminiscent of the recurrent connectivity within a cortical layer or neuronal culture~\figref{fig:weight_contrib}A.
We independently draw the strength of each connection from a log-normal distribution --- see Method \ref{subsubsec:spatial-net} for details.
Many important properties of the connectome are already captured by the adjacency matrix and derived measures (e.g. the binary clustering coefficient \ccbin).
However, for the nodes with similar binary \ccbin, different realizations of the weight distribution will still result in distinct weighted clustering coefficients.  
How much difference the particular clustering measure generates reflects its capability to capture the variance of the node properties defined by the weights. 
We first consider the nodes with binary clustering among the smallest in the network (between $]1\%--2\%]$, the 2nd percentile of all \ccbin{} values in the network, to avoid nodes with null binary clustering), median ($]49\% - 50\%]$) and highest ($]99\% - 100\%]$) and investigate which percentiles of the weighted clusterings will correspond to the same nodes depending on the selected definition.  
We observe that for the median percentile of \ccbin, all weighted methods generate large variability.
However, for the highest and smallest non-null \ccbin, the variance of the weighted ranks is much larger for the fully weighted methods.
In contrast, the hybrid methods follow the binary ranks more closely --- \figref{fig:weight_contrib}B. 

We capture the full range of \ccbin{} and characterize the variability in the weighted definitions by considering the nodes binned by \ccbin{} in bins of 0.05 and generating 20 networks to improve the statistics. 
Within each bin, we consider the distribution of the observed \ccwei. 
We compute the normalized \ccwei{} spread as an interquartile range (difference of the value at 75th percentile and 25th percentile) normalized by the median of values within the bin for a fair comparison (as the values generated by different distributions systematically vary).
For all values of binary clustering, the variability of fully-weighted measures is much larger than the hybrid measures.
Particularly, Barrat's definition closely follows \ccbin{} without large variability, the continuous definition generates around $25\%$ more variability than Onnela's, and Zhang's goes up to $65\%$ more than Onnela's --- \figref{fig:weight_contrib}C.

We showed that, in general, fully weighted methods are significantly more sensitive to the weights-related features as they separate nodes that have similar binary clustering more efficiently based on the weights' contribution. 
In the following sections, we illustrate how this increased sensitivity impacts the analysis of real connectomics data.

\subsubsection{Fully-weighted methods overcome distortions from spurious connection more reliably than thresholding}

\begin{figure*}[ht]
	\includegraphics[width=\textwidth]{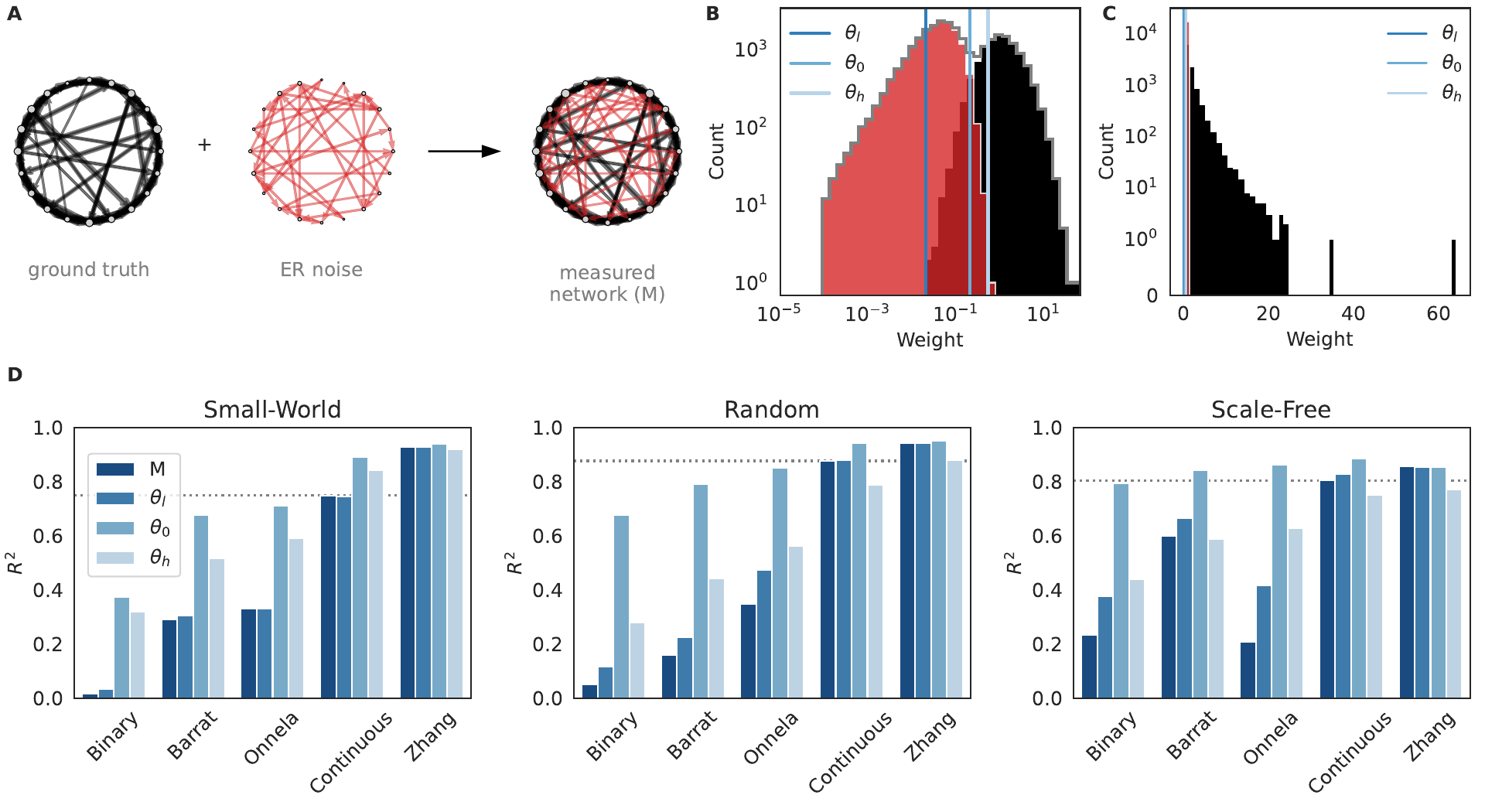}
	\caption{
		Thresholding for the removal of spurious edges strongly affect hybrid, but not fully-weighted clustering measures.
		\textbf{A}. The measured network (M) is a combination of the ground truth (black) and an Erd\H{o}s-R\'{e}nyi noise (red).
		\textbf{B}. The distribution of true (black, log-normal) and noisy (red, shifted exponential) weights have a non-zero overlap.
		Thresholds: low: $\theta_l$ - minimum of ground-truth weight, optimal  $\theta_o$, and high $\theta_h$ - maximal weight of the noise network.
		\textbf{C}. Same as B but in linear scale.
		\textbf{D}. Coefficient of determination $R^2$ (blue) measuring the match between the thresholded and ground-truth total clustering values. 
		The dotted line marks the minimal correlation for the measured network with a fully-weighted method, to be compared with the best values of the binary and hybrid methods.}
	\label{fig:thresholding_comparison}
\end{figure*}

Most networks in neuroscience, particularly on the macro and mesoscales, are inferred or derived from indirect measurements.
As a result, connectomics data is often corrupted by a large amount of noise that is unlikely to disappear soon~\cite{thomas2014anatomical}. 
A previous study~\cite{Zalesky.2016}, using non-weighted measures, demonstrated that erroneously included connections (false positives) had a higher impact on network clustering than erroneously omitted ones (false negatives). 
Therefore, the authors concluded that it is preferable to use high thresholds to discard as many spurious edges as possible.
We show that this result does not hold for the weighted clustering coefficient and that the decision about the measure to use is more critical than correct thresholding ---
~\figref{fig:thresholding_comparison}.

We test sensitivity to noise on three characteristic network models, assuming that the spurious connections will generally have smaller weights. 
We consider three ground-truth networks: random, small-world, and scale-free, with weights drawn from a log-normal distribution as was observed in the statistics of neuronal connections~\cite{buzsaki2014log, loewenstein2011multiplicative}. 
We add probability $p_\text{spur} = 0.017$ of observing a spurious edge between two unconnected nodes, the weights of spurious edges are drawn from a shifted exponential distribution starting at $10^{-4}$ with a characteristic scale of $1.5$. This way, spurious and real weight distributions overlap and cannot be separated using a single threshold, \figref{fig:thresholding_comparison}A--C.

The implicit assumption of thresholding is that a measure of interest will remain stable and closely aligned with the ground truth upon its application to the optimally thresholded network.
Because in our example, the real and spurious weight distributions are known, we can find the optimal thresholding value $\theta_0$ that enables the removal of most spurious edges while preserving most of the real edges, \figref{fig:thresholding_comparison}B.
Additionally, we define a strict threshold $\theta_h$, corresponding to the removal of all spurious edges at the cost of an increase in false negatives, and a lenient threshold, $\theta_l$, where all real edges are preserved at the cost of an increase in false positives. 
We also consider the whole measured network (M) obtained when setting the threshold at 0.
Using these thresholds, we obtain four different networks and can now compare the clustering coefficients in them to the ground truth. 
To this end, we regress the local clustering coefficients \ccwei{} in a thresholded or complete network on the ground-truth clustering coefficient obtained by the same method in the noise-free network. 

The most striking result from this comparison is that, regardless of the threshold, fully weighted measures provide a notably higher correlation with the ground truth
compared to binary and hybrid definitions --- \figref{fig:thresholding_comparison}D.
In fact, the best correlation coefficients (using the optimal threshold) for the binary clustering are lower those obtained by applying directly the fully-weighted methods on the measured graph.
Even for hybrid measures, their optimal results are at best similar or only marginally higher than the correlation coefficient of fully-weighted methods on the measured graph, and the correlation can drop very quickly as soon as one deviates from the ``optimal'' threshold. 

Correctly thresholding the network led to notable improvements for all methods, improving the correlation to the ground truth.
However, finding the precise threshold requires knowledge of the real and spurious weight distribution, and spurious links strongly affect the local clustering for binary and hybrid methods as soon as one deviates from the optimal threshold.
Furthermore, contrary to what happens in binary networks, higher thresholds may decrease the correlation to the ground truth depending on the graph and method.
On the other hand, considering the entire network but using fully weighted measures allows us to obtain a nearly perfect inference of the clustering ground-truth network.  
The reason for this lies in the methods' mathematical properties, ensuring that spurious low-weight edges have little influence on the local clustering of a node. 

For neuroscience applications, we expect that the contribution of a triangle to the weighted clustering coefficient of a neuron should go to zero if any of the weights in the triangle goes to zero.
Because the Zhang--Horvath definition does not fulfill this condition and has been shown to be less sensitive to some weight-encoded features, we will default to the continuous clustering when using a fully-weighted measure in some of the following sections (see \cite{fardetWeightedDirectedClustering2021} for detailed discussion and examples).
However, it is worth noting that most of the results still hold for the Zhang--Horvath definition and that it may be preferred in some cases due to its higher robustness to spurious edges.

\subsection{Connectomics Data}

\subsubsection{Directed clustering patterns in brain networks}

\begin{figure*}
	\includegraphics[width=\textwidth]{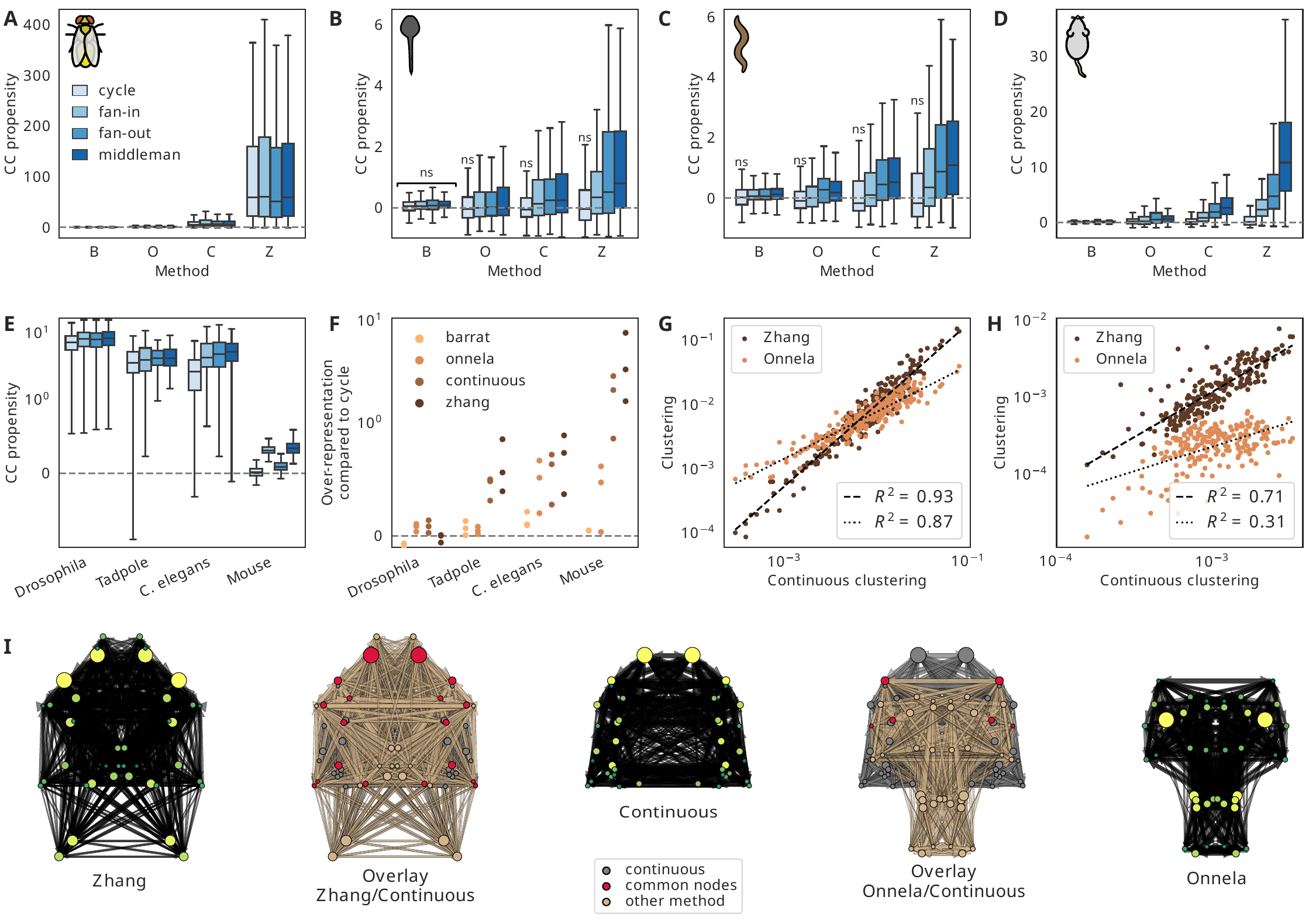}
	\caption{
		Expression of directed clustering motifs can vary significantly between clustering methods and connectomes.
		\textbf{A--D}. Clustering propensity of four connectomes for each directed motif and each clustering definition.
		The values are compared to the average of 10 networks with the same binary structure but shuffled weights (see Methods \ref{methods:analysis}, for details).
		The clustering definitions used are Barrat's, ``B'', and Onnela's, ``O'' (hybrid measures) as well as continuous, ``C'', and Zhang--Horvath's, ``Z'' (fully-weighted measures).
		\textbf{A}. The central brain of drosophila displays a large increase in all directed motifs according to the fully-weighted measures.
		\textbf{B}. The \textit{Ciona intestinalis} tadpole connectome shows almost no effect of weight structure except for a slight increase in middleman motifs and a slight decrease of cycles with the Zhang--Horvath definition.
		\textbf{C}. In the \textit{C. elegans} connectome, weights contribute to a reinforcement of middleman and fan-out patterns and a slight reduction of cycles according to the fully-weighted measures.
		\textbf{D}. Weight distribution in the mouse mesoscale connectome contributes to a notable increase of middleman, fan-in and fan-out motifs but not to cycles, according to the fully-weighted methods.
		\textbf{E}. Using the continuous clustering, comparison between the actual expression levels of the motifs and the expected value in Erd\H{o}s-Rényi graphs (with the same number of edges and the same set of weights) reveals that, except for cycle motifs in the mouse connectome, all clustering motifs are over-expressed.
		\textbf{F}. Level of expression for the middleman, fan-in, and fan-out motifs compared to cycles according to the four clustering definitions.
		Each dot represents the ratio of the median value of a motif compared to the median cycle clustering.
		The fully-weighted methods detect significant over-expression of the three motifs compared to cycles for all connectomes except drosophila.
		\textbf{G, H}. In \textit{C. elegans}, correlations are significant between the continuous clustering and the Zhang--Horvath and Onnela distributions and strong correlations are also observed in the tadpole and drosophila; however, much weaker correlations are observed for the mouse brain.
		\textbf{I}. As expected from the low correlations seen in H, the subnetworks formed by the most clustered areas in the mouse brain differ strongly depending on the method used.
		Node size express total-degree, colors give the total clustering values, from high (yellow) to low (dark green).
		For the overlays, nodes and edges from the continuous subnetwork are in grey while Zhang/Onnela are in orange.
		Nodes present in both subnetworks are marked in red.
	}
	\label{fig:brain-motifs}
\end{figure*}

Having demonstrated the advantages of fully weighted directed methods in artificial networks, we proceed to test how various clustering definitions perform on actual connectomics data. Specifically, we use networks derived from the connectomes of four different species: mouse \cite{Oh.2014},  \emph{Drosophila melanogaster} (drosophila, \cite{Scheffer.2020}), \emph{C. elegans} \cite{Cook.2019}, and tadpole of \emph{C. intestinalis} (tadpole \cite{Ryan2016}). For these networks, we study the ability of different methods to extract information from the weighted network. In the following, we will investigate which of the directed clustering patterns (cycle, fan-in, fan-out, and middleman) are over- or underrepresented in biological networks compared to random weight distribution.

We are particularly interested in the impact of weight distribution beyond the information already contained in the adjacency matrix.  
To access it, we compare the weighted clustering $CC_w$ (using each of the four candidate methods) of the original connectomes to a null model with the same binary connectivity but where all weights have been randomly shuffled. Whereas Barrat's and Onnela's definitions do not capture any significant difference in weighted connectivity motifs over binary, the fully weighted definitions detect a significant over-representation of some directed patterns, notably for middlemen but also for fan-in and fan-out. These results are consistent across species and occur both at the microscale (\emph{C. elegans}, tadpole) and at the mesoscale (neuropills of fly central complex and mouse brain) \figref{fig:brain-motifs}A--D.

The relative contribution of weights (compared to the contribution of the binary network topology) varies greatly from connectome to connectome.
Sparse networks, namely the microscale ones and the mesoscale fly central complex, are quite strongly structured, whereas the dense mesoscale mouse brain displays a much weaker binary structure, as can be seen from \figref{fig:brain-motifs}F.
Because of this, the correlation between $CC_w$ methods is much higher in structured networks such as \emph{C. elegans} --- \figref{fig:brain-motifs}E --- than in the mouse brain --- \figref{fig:brain-motifs}G.

The difference between middleman/fan-in/fan-out and circular motifs comes from both weight-specific and structural contributions in some connectomes. Both \emph{C. elegans} and the mouse brain display significant differences between these two types of clustering patterns. This effect is less significant for the tadpole and minimal in the fly central complex, where no significant difference can be seen between the motifs' expression levels.

Finally, depending on the method used, the nodes displaying the highest clustering values can differ significantly --- \figref{fig:brain-motifs} H-I. This is especially visible for the mouse brain where top-clustered areas change almost completely between Onnela's and the continuous definition while they display a higher number of common nodes with the Zhang--Horvath definition.

In summary, our analysis reveals that directed motifs are under- or over-represented in brain networks and that the over-representation of weighted patterns associated with redundant information transfer is visible across species and scales.
Moreover, to see the specific contribution of connection weights to node clustering, it is necessary to use a fully-weighted method.

\begin{figure*}
	\includegraphics[width=\textwidth]{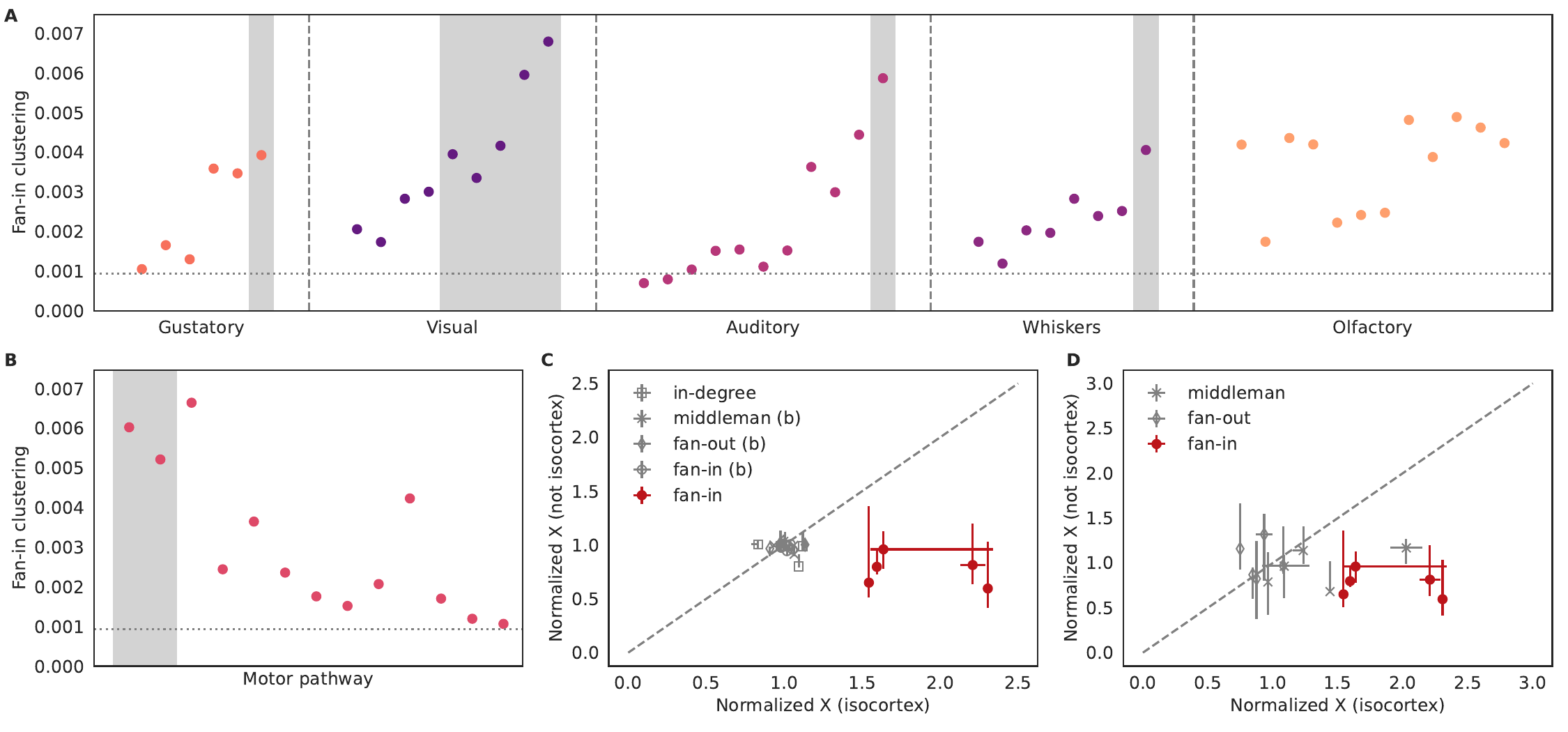}
	\caption{
		The continuous weighted fan-in clustering coefficient correlates with the functional organisation of most sensory and motor pathways. 
		\textbf{A}. Evolution of the fan-in clustering along several sensory pathways (see the mouse connectome entry in the Methods section (\ref{subsec:method-connectomes}) for a detailed list of the selected areas for each pathway).
		\textbf{B}. Evolution of the fan-in clustering along the motor pathway.
		In both A and B, areas belonging to the isocortex are highlighted in grey.
		The olfactory pathway is the only one that does not involve the isocortex nor displays a clear trend in the clustering coefficient.
		\textbf{C}. Compared to the binary measures, marked by (b), the weighted fan-in clustering efficiently discriminates the isocortical areas from the rest of the pathway areas: isocortical areas (given by the x-axis) have larger-than-average clustering.
		\textbf{D}. Even among weighted clustering patterns, fan-in is the only motif that displays a consistent trend along the pathways, with all values significantly below the identity line (dashed).
		To make them comparable, all properties $X$ were normalized by their average value $\langle X \rangle$ in the network, and the error bars denote the interquartile of the distribution.
	}
	\label{fig:mouse-pathways}
\end{figure*}

\subsubsection{Fully-weighted clustering methods reveal hierarchical structures along the mouse cortical pathways}

To better demonstrate the impact of choosing between hybrid and fully weighted directed clustering methods, we highlight how the fully weighted approach is more effective at revealing significant, large-scale patterns in connectomics data, patterns which the hybrid method fails to discern. 
To this end, we use the mesoscale mouse connectome \cite{Oh.2014} and investigate different sensory pathways. 

There is a natural hierarchy within the sensory and motor areas of the mammalian brain. This hierarchy will likely manifest in the interactions between these areas, potentially reflected in connectivity metrics such as clustering coefficient. Our hypothesis is that higher-level brain areas integrate information from multiple levels of this hierarchy. Therefore, our focus is on investigating the behaviour of fan-in clustering, as it could provide insights into how these upper-tier regions receive, process and consolidate information from various interconnected sources.

Indeed, using fully-weighted clustering methods to estimate the fan-in clustering, we observe that  (with the exception of the olfactory pathway, where no pattern is discernible), the clustering steadily increases as we move to higher areas in the sensory (\figref{fig:mouse-pathways}A) and motor pathways (\figref{fig:mouse-pathways}B). 

The significantly above-average fan-in clustering at the high-level sensory and motor processing areas illustrates the interplay between structured connectivity (in this case, fan-in clustering) and function (the processing of more complex information taking place in the higher cortical areas). The exact mechanisms by which the connectivity structures contribute to the specific functions remain unclear, but we can at least establish a clear correlation that, as far as we are aware, has not been observed before.

Interestingly, the fan-in clustering is the only one among multiple binary (\figref{fig:mouse-pathways}C) and weighted (\figref{fig:mouse-pathways}D) metrics that correlates well with the cortical hierarchy. In particular, it allows us to effectively discriminate isocortical areas from the other pathway areas via the higher clustering of the former. None of the other clustering metrics presents such a correlation (\supfigref{fig:supp-mouse-measures-comp}).

Crucially, we see that this link between structure and function can be uncovered only via the fully-weighted, directed methods: while both the continuous and Zhang's definitions capture the gradual increase of the weighted fan-in clustering along the different pathways well, the hybrid methods fail to establish any correlation  (\supfigref{fig:supp-mouse-fan-in-comp}).

\subsubsection{On the small-worldness of brain networks}
\begin{figure*}
	\includegraphics[width=\textwidth]{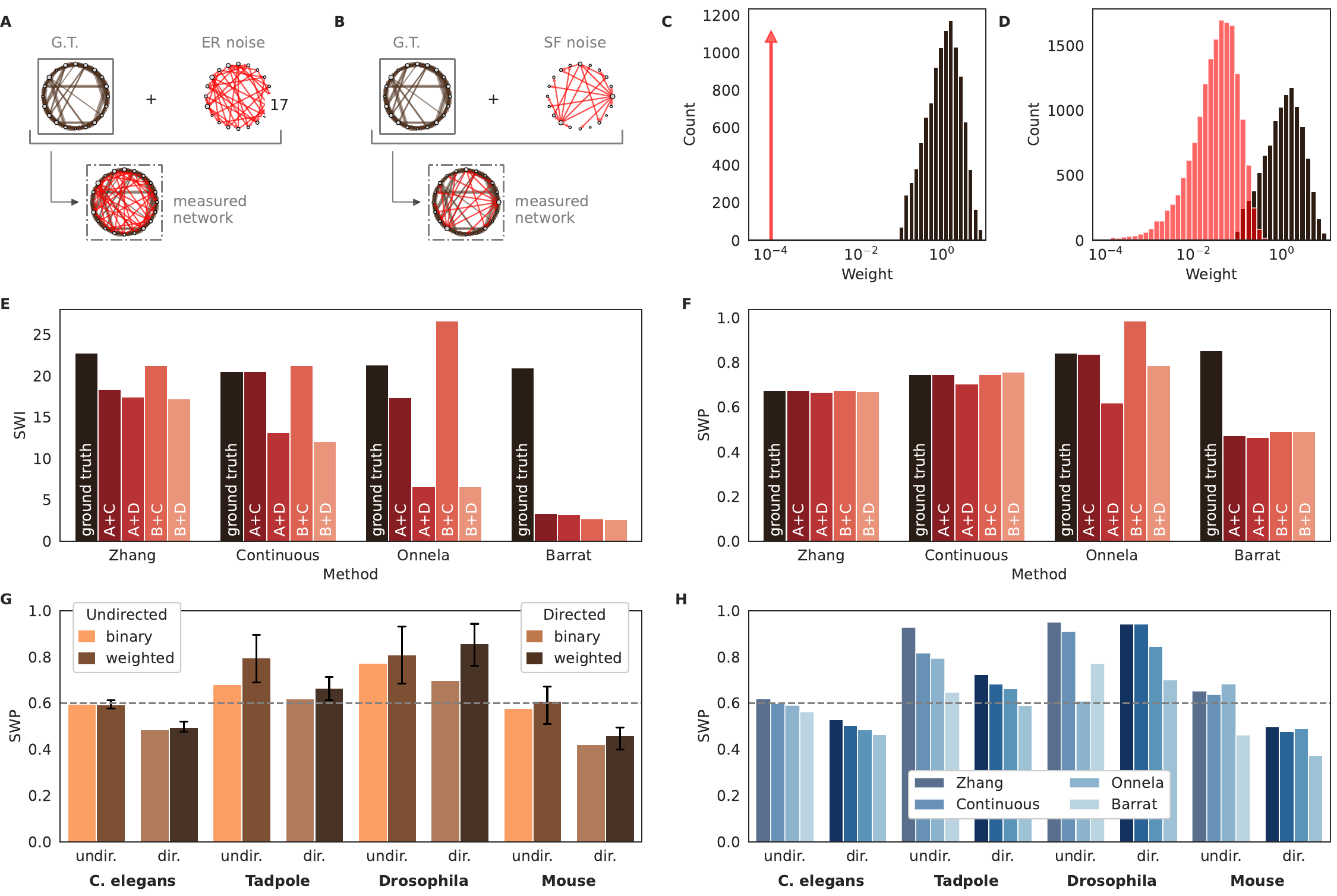}
	\caption{
		Small-worldness measures vary and display different robustness to noise depending on the clustering method used.
		Most connectomes do not register as small-worlds according to the directed-weighted small-world propensity (SWP).		
		\textbf{A--D}. A ``measured network'' can be represented as the union of a ``ground truth'' (G.T.) --- here a Watts--Strogatz network in dark brown --- and spurious small-weight connections (``noise'' graphs with: A random, or B scale-free connectivity, in red).
		We assess the influence of the weight distribution of spurious connections (red) by checking weights that are: \textbf{C} all equal and small or \textbf{D} following an exponential distribution and overlapping with the real weights (dark brown).
		\textbf{E}. Early measures of small-worldness, such as the small world index (SWI) display high variability depending on the noise and the method.
		\textbf{F}. The more recent small-world propensity (SWP) notably reduces this variability, especially for fully-weighted measures, though hybrid measures still present marked fluctuations.
		\textbf{G}. Though the connectomes often display small-world tendencies (proposed as SWP $>$ 0.6 in \cite{Muldoon2016}) for their undirected binary (grey) or weighted (light steelblue) versions, the situation can change drastically for the directed versions (dark grey and steelblue).
		Both \textit{C. elegans} and the mouse connectome have relatively low small-world propensity whereas the tadpole and drosophila display higher values, registering as small-world.
		As all connectomes are weakly connected, the undirected SWP always uses the full network; however, only the mouse connectome is strongly connected so, for directed SWP in the other connectomes we use the largest strongly connected component (SCC) which contain between 85 and 99\% of the neurons in the original networks.
	}
	\label{fig:swp}
\end{figure*}

We finally extend the measures of small-worldness to weighted-directed versions using the directed extensions of the clustering coefficient and directed null-model graphs for the small-world propensity, as detailed in the Methods section \ref{methods-swp}. In this section, for the directed clustering coefficient, we take the total clustering coefficient without considering individual motifs. 

To test how small-worldness measures are affected by the underlying clustering definitions, we construct a Watts-Stogatz network \ref{methods:WS_network_details} to which we add spurious connections with small-weights. The spurious connections are drawn from a network with either random (\figref{fig:swp}A) or scale-free connectivity (\figref{fig:swp}B) and the connection weights can be either all equal to a small value (\figref{fig:swp}C) or drawn from an exponential distribution (\figref{fig:swp}D), leading to four different combinations of spurious connectivity.

We begin by examining how the small word index (SWI) changes for the different types of spurious connections according to the underlying clustering definition. We find that while a strong divergence from the ground truth is visible regardless of the clustering definition used, hybrid definitions fare significantly worse compared to fully-weighted ones (\figref{fig:swp} E). We then examine a more recent and reliable metric of small-worldness, the small-world propensity (SWP) \cite{Muldoon2016}, and we observe that it is significantly more robust to noise regardless of the underlying clustering definition. Still, we clearly see that using the Zhang--Horvath or continuous definitions leads to more reliable results than using hybrid methods (\figref{fig:swp}F).

Finally, we find that using directed and weighted small-world measures on connectomes reveals that small-worldness is not a universal feature of neuronal networks per se. 
Following Muldoon \emph{et al.} \cite{Muldoon2016} improvements with the introduction of small-world propensity, it is still generally accepted in the neuroscience community that the ``small-world'' label should be attributed to a network when its small-world propensity goes over a somewhat arbitrary threshold of SWP = 0.6.
However, we see that while the undirected versions of the connectomes (both weighted and unweighted) satisfy this definition of small-worldness, the equivalent directed versions have a significantly lower SWP in three out of the four connectomes (with the exception of the fruit fly).
The resulting SWP is significantly below $0.6$ for \emph{C. elegans} and below $0.5$ for the mouse brain, but remains above the threshold for the fruit fly and the tadpole of \emph{C. intestinalis} (\figref{fig:swp}G).
Thus, while the connectomes of the fruit fly and the tadpole of \emph{C. intestinalis} retain a high small-world propensity even when properly assessed as weighted directed networks, the connectomes of \emph{C. elegans} and the mouse cannot be considered small-world because of their low clustering levels and long path length --- see \supfigref{fig:supp_swp}.
These results are consistent for all clustering methods, although hybrid methods tend to estimate lower SWP across connectomes (\figref{fig:swp}H).


\section{Discussion}

Contrary to what happens for binary clustering, we showed that when using weighted measures, false negatives (removal of real edges) can lead to greater errors than false positives (presence of spurious edges) depending on the network structure.

As the problem of finding a sensible threshold for a given network remains unsolved, we suggest a different approach to deal with potentially spurious low-weight edges. 
To properly assess strong recurrent interactions in weighted networks, one needs a fully weighted definition of clustering that is not influenced by small and spurious connections. 
Such a measure might eliminate the need for thresholding and enable analysis to be performed directly on a measured network, even if it contains many spurious edges with low weights.

Our analysis of connectomes enabled us to highlight how binary structure and weight correlations contribute to reinforcing redundant information transfer via middleman, fan-in, and fan-out motifs (\figref{fig:brain-motifs}).
In the mouse connectome, fan-in clustering strongly correlates with the organization of most sensory and motor pathways (\figref{fig:mouse-pathways}).
As such neurons need to reliably integrate large amounts of information, these large clustering values reinforce the hypothesis that redundant and clustered structures play a significant functional role in such circuits.
Indeed, as one progresses along sensory pathways, the areas tend to address more and more complex signals, potentially via a combination of the results from the previous areas and the additional synergistic information left in the initial signal, which would require such fan-in motifs.
In the mouse, these areas mostly belong to the isocortex except for notable outliers such as caudoputamen (CP), which has high clustering values due to its very large size and, therefore, large associated weights, and the inferior and superior colliculus (IC/SC) that are involved in multi-sensory integration \cite{grutersSoundsMultisensoryOther2012,itoMouseSuperiorColliculus2018} and are therefore expected to have high fan-in values.

Finally, regarding small-worldness, one can see a much more complex picture than the previously proposed universal small-worldness of brain networks.
Indeed, both at the micro- and mesoscale levels, we find that some connectomes are far from being small-world when compared to the commonly used threshold. Notably, we identify low clustering values and high path lengths for \textit{C. elegans}, and the mouse.
Furthermore, most connectomes are not strongly connected, meaning that not all nodes can reach all other nodes (which goes against the concept of small-worldness); this corresponds to 9\% of the nodes for \textit{C. elegans} and \textit{C. intestinalis}, and 3\% for drosophila.
Small-worldness has stirred controversy in neuroscience, and though many of the technical issues raised a few years ago \cite{papo_beware_2016} have been partly addressed thanks to weighted analysis methods and the introduction of the SWP, the fundamental question of its relevance remains.
Indeed, as discussed in \cite{papo_beware_2016, avena-koenigsberger_spectrum_2019}, the relevance of shortest paths for communication in brain networks is questionable.
Furthermore, the notion of clustering associated with small-worldness ignores the specificity of directed patterns that play a crucial in identifying specific populations and processing functions.

\subsection*{Open questions and limitations}

Among all sensory pathways in the mouse connectome, the olfactory pathway was the only one that did not display any correlation between the pathway hierarchy and the fan-in clustering values.
The fact that this is the only pathway where all nodes belong to the cerebral cortex (and mostly to the same region, the olfactory areas) is probably not a coincidence, but it is unclear why this particular pathway would behave differently from the other sensory pathways.


The immature connectome that we investigated, for the tadpole of \textit{C. intestinalis}, was the only chemical network where no significant influence of connection strength was visible, is also of great interest.
Further work on the connectome of a single animal at different developmental stages would bring significant benefits to assess whether clustering and small-world properties vary or remain stable throughout development and how this could relate to functional and behavioural features.

\textit{C. elegans} is an interesting example of small-worldness because it consists of two complementary networks: one composed of chemical synapses has low SWP (cf. \figref{fig:swp} G-H) and another, undirected and with high SWP, consists of electrical (or gap) junctions that directly couple the membrane potentials or connected cells.
However, these networks correspond to very different types of interactions, and it is therefore unclear whether we can simply combine them together to say that the overall connectome of \textit{C. elegans} displays small-world properties.


\section{Methods}

All networks were generated and analysed using version 2.7.1 of the NNGT library \cite{NNGT}.

\subsection{Connectomes}

\label{subsec:method-connectomes}

This part describes the datasets that were used to generate the connectomes used in this study.
The code to build the actual networks from the data is available upon request and will be made public once the article is published.

\tocless{\subsubsection*{\emph{Caenorhabditis elegans}}}

The worm connectome was taken from the WormWiring website\footnote{\href{https://wormwiring.org/pages/adjacency.html}{https://wormwiring.org/pages/adjacency.html}} as the network of chemical synapses from the hermaphrodite \textit{C. elegans} (\cite{Cook.2019} SI4 and SI5).
It was mapped by Cook \textit{et al.} using electron microscopy.
We ignore self-loops present in the original data and obtain a network of 302 neurons and 3,709 directed chemical synapses.
This corresponds to a network density of \SI{4.1}{\percent}. 
Weights are discrete, ranging from 1 to 75, and correspond to the total amount of physical connectivity between neurons, including the number and size of synapses.

The strongly-connected component contains 275 neurons.

\tocless{\subsubsection*{Tadpole of \emph{Ciona intestinalis}}}

The tadpole microscale connectome was obtained from \cite{Ryan2016}, from which we investigated the chemical synapses between all 222 cells.

The data is available from the supplementary data 1 from Figure 6\footnote{\href{https://elifesciences.org/articles/16962/figures\#SD6-data}{https://elifesciences.org/articles/16962/figures\#SD6-data}}.

The network contains 222 cells and 3,105 edges with weights, reflecting the size of the synaptic connections in micrometer, with a range from 0.06 to 16.623 for neurons and to more than 114 for connections to muscle nodes.

The strongly-connected component contains 196 neurons and 2,972 edges with weights from 0.06 to 16.623.

\tocless{\subsubsection*{\emph{Drosophila melanogaster}}}

The fruit fly connectome was taken from the hemibrain connectome v1.2 which contains a dense reconstruction of the fruit fly's central brain \cite{Scheffer.2020}. 
The connectome was reconstructed using electron microscopy and a combination of machine learning algorithms and manual labor for neuron and synapse detection.
We accessed the neuPrint database \footnote{\href{https://neuprint.janelia.org/}{https://neuprint.janelia.org/}} through its Python API \footnote{\href{https://github.com/connectome-neuprint/neuprint-python}{https://github.com/connectome-neuprint/neuprint-python}}.

From the initial 97,900 neurons present in the dataset, we kept the 85,189 that had both in- and out-connections.
Consistently with the way neurons were grouped in \cite{Scheffer.2020} and \cite{Hulse2020}, we associated them by their two main input and output neuropils (potentially only one of each if a single input or output neuropil was associated to the neuron).
Considering all synapses, this gave us a mesoscale connectome consisting of 2,349 nodes containing from 1 to 18,570 neurons each and 328,548 edges.
The weights associated to the connections span 6 orders of magnitude, from 1 to 2,931,945.

The strongly connected component contains 2,347 nodes.

If we follow instead the methodology from and keep only strong connections (at least 3 synapses between the neurons), this leads 52,668 neurons with in- and out-connections that were coalesced into a network of 2,356 nodes containing from 1 to 9,565 neurons each and connected via 144,368 edges.
The edge weights span 6 orders of magnitude, from 3 to 2,250,066.

The strongly-connected component contains 2,338 nodes.

\tocless{\subsubsection*{Mouse}}

The mesoscale network from the mouse brain was obtained from the NeuroData\footnote{\href{https://neurodata.io/project/connectomes/}{https://neurodata.io/project/connectomes/}} file Mouse\_brain\_1, obtained from the work of \cite{Oh.2014}.

We regenerated the whole brain from this by duplicating each node into its left/right hemisphere instance and generating the associated ipsi- and contra-lateral connections.
In the end, this led to a network of 426 nodes and 65,464 edges with weights spanning more than 16 orders of magnitude, ranging from almost zero to slightly more than 20.

All nodes are contained within a single strongly-connected component.

For the pathways detailed on \figref{fig:mouse-pathways}, the associated areas as detailed below, in the same order as in the figure.
Information about the pathways were obtained from \cite{Oh.2014} (Extended Data Figure 10) and \cite{pailthorpeNetworkAnalysisMesoscale2019} as well as from pathway-specific studies.\\

For the gustatory pathway \cite{carletonCodingMammalianGustatory2010}, the areas are: Nucleus of the solitary tract (NTC), Parabrachial nucleus (PB), Ventral posteromedial nucleus of the thalamus (parvicellular part, VPMpc), Lateral hypothalamic area (LHA), Basolateral amygdalar nucleus (BLA), Gustatory areas (GU).\\
For the visual pathway, the areas are: Dorsal part of the lateral geniculate complex (LGd), Ventral part of the lateral geniculate complex (LGv), Superior colliculus, sensory related (SCs), Lateral posterior nucleus of the thalamus (LP), Lateral visual area (VISl), Anteromedial visual area (VISam), Anterolateral visual area (VISal), Primary visual area (VISp).\\
For the auditory pathway \cite{dibonitoCellularMolecularUnderpinnings2017}, the areas are: Dorsal cochlear nucleus (DCO), Ventral cochlear nucleus (VCO), Superior olivary complex (SOC), Nucleus of the lateral lemniscus (NLL), Medial geniculate complex (ventral part, MGv), Medial geniculate complex (medial part, MGm), Medial geniculate complex (dorsal part, MGd), Inferior colliculus (external nucleus, ICe), Inferior colliculus (dorsal nucleus, ICd), Inferior colliculus (central nucleus, ICc), Primary auditory area (AUDp).\\
For the whisker sensory pathway, the areas are: Spinal nucleus of the trigeminal (oral part, SPVO), Spinal nucleus of the trigeminal (caudal part, SPVC), Spinal nucleus of the trigeminal (interpolar part, SPVI), Principal sensory nucleus of the trigeminal (PSV), Superior colliculus, sensory related (SCs), Posterior complex of the thalamus (PO), Ventral posteromedial nucleus of the thalamus (VPM), Primary somatosensory area (barrel field, SSp-bfd).\\
For the olfactory pathway \cite{decastroWiringOlfactionCellular2009,moriOlfactoryConsciousnessGamma2013}, the areas are: Main olfactory bulb (MOB), Accessory olfactory bulb (AOB), Anterior olfactory nucleus (AON), Olfactory tubercle (OT), Nucleus of the lateral olfactory tract (NLOT), Taenia tecta (TT), Cortical amygdalar area (anterior part, COAa), Cortical amygdalar area (posterior part, COAp), Piriform-amygdalar area (PAA), Piriform area (PIR), Postpiriform transition area (TR), Entorhinal area (lateral part, ENTl).\\
For the motor pathway \cite{jeongComparativeThreedimensionalConnectome2016}, the areas are: Primary motor area (MOp), Secondary motor area (MOs), Caudoputamen (CP), Globus pallidus (internal segment, GPi), Globus pallidus (external segment, GPe), Ventral anterior-lateral complex of the thalamus (VAL), Parafascicular nucleus (PF), Ventral medial nucleus of the thalamus (VM), Mediodorsal nucleus of thalamus (MD), Superior colliculus (motor related, SCm), Subthalamic nucleus (STN), Pontine gray (PG), Nucleus raphe magnus (RM).

\subsection{Nomenclature}

For all definitions, we use the following notation:
$A$ is the adjacency matrix of a graph. 
$W = \{w_{ij}\}$ is the normalized weight matrix, obtained from the original weight matrix $\tilde{W} = \{\tilde{w}_{ij}\}$ by $w_{ij} = \tilde{w}_{ij} / \max_{i,j}\left(\tilde{w}_{ij}\right)$. 
$d_i$ is the degree of node $i$ and $s_i = \sum_j w_{ij}$ is its strength.
$W^{\left[ \alpha \right]} = { w_{ij}^\alpha }$ and $s^{\left[ \alpha \right]}_i = \sum_{j \sim i} w_{ij}^\alpha$ are the fractional weight matrix and strength for any $\alpha \in \mathbb{R}$.

For the directed definitions, $d_{i, \text{in}}$, $d_{i, \text{out}}$ and $d_{i, \text{tot}} = d_{i, \text{in}} + d_{i, \text{out}}$ denote the in-, out-, and the total-degree of node $i$. Similarly, $s_{i, \text{in}}$, $s_{i, \text{out}}$ and $s_{i, \text{tot}} = s_{i, \text{in}} + s_{i, \text{out}}$ denote the in-, out-, and the total-strength of node $i$.

The number of reciprocal edges between $i$ and its neighbors (i.e. the number of nodes j for which both an edge $i \rightarrow j$ and an edge $j \rightarrow i$ exist) is denoted by $d^{\leftrightarrow}_i$. 
The reciprocal strength of $i$ is denoted by $s_{i, \leftrightarrow}$ and can either be defined via the arithmetic mean of reciprocal weights as in \cite{Clemente.2018}, for the Barrat's definition $s^B_{i, \leftrightarrow} = \sum_{j \neq} \frac{w_{ij} + w_{ji}}{2}$, or via their geometric mean for the Zhang--Horvath and continuous definitions, $s_{i, \leftrightarrow} = \sum_{j \neq i} \sqrt{w_{ij} w_{ji}}$. 
Details are provided if the reciprocal strength is used in a definition of clustering coefficients.

\subsection{Weighted directed clustering definitions}
\label{methods-clustering}

Weighted directed clustering measures use the edge weights to quantify the importance of each triangle, called triangle ``intensity''.
Instead of counting the number of triangles $\Delta_{ijk}$ that a node $i$ participates in, they sum the intensities $I_{\Delta ijk}$ of these triangles.
Similarly, they can use the same methodology for to quantify triplets via the triplet intensity $I_{T ijk}$.
Differences between weighted measures thus lay in the way they define triangle and triplet intensity as the generic definition of the clustering coefficient for a definition $D$ and a motif $m$ is:
\begin{equation}
	C^{D,m}_i = \frac{I^{D,m}_{\Delta,i}}{I^{D,m}_{T,i}} = \frac{\sum_{j\neq k} I^{D,m}_{\Delta ijk}}{\sum_{j\neq k} I^{D,m}_{T ijk}}
\end{equation}

We separate the weighted measures into two classes:
\begin{itemize}
	\item Hybrid definitions, that combine weighted properties (from the weighted connectivity matrix) with binary properties that come from the adjacency matrix (typically node degrees or average weights).
	\item Fully-weighted definitions, that do not include any binary properties and only use the weighted connectivity matrix.
\end{itemize}

The two hybrid definitions were introduced by Barrat et al. \cite{Barrat.2004} and Onnela et al. \cite{Onnela.2005}, then extended to directed networks by \cite{Clemente.2018} and \cite{Fagiolo.2007} respectively.
The fully weighted definitions include that from Zhang and Horvath \cite{Zhang.2005}, originally introduced for undirected networks, then extended to directed networks and the continuous definition \cite{fardetWeightedDirectedClustering2021}.

\begin{table*}[t]
	\centering
	\begin{ruledtabular}
		\begin{tabular}{r c c c c}
			 & \textbf{Cycle} & \textbf{Fan-in} & \textbf{Fan-out} & \textbf{Middleman}\\\hline
			$\boldsymbol{I^{O,m}_{\Delta,i}}$ & $\left(W^{\left[\frac{1}{3}\right]}\right)^3_{ii}$ & $\left(W^{\left[\frac{1}{3}\right]T}\left(W^{\left[\frac{1}{3}\right]}\right)^2\right)_{ii}$ & $\left(\left(W^{\left[\frac{1}{3}\right]}\right)^2 W^{\left[\frac{1}{3}\right]T}\right)_{ii}$ & $\left(W^{\left[\frac{1}{3}\right]}W^{\left[\frac{1}{3}\right]T}W^{\left[\frac{1}{3}\right]}\right)_{ii}$\\\\
			
			$\boldsymbol{I^{O,m}_{T,i}}$ & $d_{i, \text{in}}d_{i, \text{out}} - d_{i, \leftrightarrow}$ & $d_{i, \text{in}} (d_{i, \text{in}} - 1)$ & $d_{i, \text{out}} (d_{i, \text{out}} - 1)$ & $d_{i, \text{in}}d_{i, \text{out}} - d_{i, \leftrightarrow}$\\\\
			
			$\boldsymbol{I^{B,m}_{\Delta,i}}$ & $\frac{1}{2}\left( W A^2 + \left(W A^2\right)^T \right)_{ii}$ & $\frac{1}{2} \left(W^T(A + A^T)A\right)_{ii}$ & $\frac{1}{2}\left(W(A+A^T)A^T  \right)_{ii}$ & $\frac{1}{2}\left( W A^T A + W^T A A^T\right)_{ii}$\\\\

			$\boldsymbol{I^{B,m}_{T,i}}$ & $\frac{1}{2}\left(s_{i, \text{in}}d_{i, \text{out}}  + s_{i, \text{out}}d_{i, \text{in}}\right) - s^B_{i,\leftrightarrow}$ & $s_{i, \text{in}} (d_{i, \text{in}} - 1)$ & $s_{i, \text{out}} (d_{i, \text{out}} - 1)$ & $\frac{1}{2}\left(s_{i, \text{in}}d_{i, \text{out}}  + s_{i, \text{out}}d_{i, \text{in}}\right) - s^B_{i,\leftrightarrow}$\\\\
			
			$\boldsymbol{I_{\Delta, i}^{Z,m}}$ & $\left(W\right)^3_{ii}$ & $\left(W^{T} W W\right)_{ii}$ & $\left(W W W^{T}\right)_{ii}$ & $\left(W W^{T}W \right)_{ii}$\\\\
			
			$\boldsymbol{I_{T, i}^{Z, m}}$ & $s_{i, \text{in}}s_{i, \text{out}} - s^{[2]}_{i,\leftrightarrow}$ & $s_{i, \text{in}}s_{i, \text{out}} - s^{[2]}_{i,\leftrightarrow}$ & $\left(s_{i, \text{in}}\right)^2 - s^{[2]}_{i, \text{in}}$ & $\left(s_{i, \text{out}}\right)^2 - s^{[2]}_{i, \text{out}}$\\ \\
			
			$\boldsymbol{I_{\Delta, i}^{C,m}}$ & $\left(W^{\left[\frac{2}{3}\right]}\right)^3_{ii}$ & $\left(W^{\left[\frac{2}{3}\right]T}\left(W^{\left[\frac{2}{3}\right]}\right)^2\right)_{ii}$ & $\left(\left(W^{\left[\frac{2}{3}\right]}\right)^2 W^{\left[\frac{2}{3}\right]T}\right)_{ii}$ & $\left(W^{\left[\frac{2}{3}\right]}W^{\left[\frac{2}{3}\right]T}W^{\left[\frac{2}{3}\right]}\right)_{ii}$\\\\

			$\boldsymbol{I_{T, i}^{C, m}}$ & $s^{\left[\frac{1}{2}\right]}_{i, \text{in}}s^{\left[\frac{1}{2}\right]}_{i, \text{out}} - s^\leftrightarrow_i$ & $\left(s^{\left[\frac{1}{2}\right]}_{i, \text{in}}\right)^2 - s_{i, \text{in}}$ & $\left(s^{\left[\frac{1}{2}\right]}_{i, \text{out}}\right)^2 - s_{i, \text{out}}$ & $s^{\left[\frac{1}{2}\right]}_{i, \text{in}}s^{\left[\frac{1}{2}\right]}_{i, \text{out}} - s^\leftrightarrow_i$
		\end{tabular}
	\end{ruledtabular}
	\caption{Definitions of the Barrat (B), Onnela (O), Zhang-Horvath (Z), and continuous (C) intensities for each partial mode pattern in directed graph.
		The triplet intensity of Onnela's method is actually the same as the binary triplet count $n^m_{T,i}$. For Barrat's triplet intensity, the reciprocal strength has been defined in \citep{Clemente.2018} as $s^B_{i, \leftrightarrow} = \sum_{i\neq j} \frac{1}{2} (w_{ij} + w_{ji})$.}
	\label{tab:partial-clustering_others}
\end{table*}

\subsection{Graph generation and shuffling}

\label{methods:analysis}

\tocless{\subsubsection{Spatial network (\figref{fig:weight_contrib})}}

\label{subsubsec:spatial-net}

The spatial network is obtained by generating 1000 nodes uniformly, at random, within a disk of radius 300.
From each node $i$, an outgoing edge to each other node $j$ in the graph is then generated with probability

\begin{equation}
	p(\mathbf{r}_i, \mathbf{r}_i) = \max\left(1 - \frac{\Vert \mathbf{r}_i - \mathbf{r}_j \Vert}{\lambda}, 0\right)
\end{equation}

On the figure, a value of $\lambda = 50$ is used.

The weight distribution is sampled from a log-normal distribution with a location of 0 and a scale of 1 associated to the underlying normal distribution. This leads to the resulting distribution having a median value of 1, a mean of $e^{\frac{1}{2}}$, and 95\% of the weight values being in the [0.14, 7] interval.

The function used in the code is: \\ \href{https://nngt.readthedocs.io/en/stable/modules/generation.html#nngt.generation.distance_rule}{\texttt{nngt.generation.distance\_rule}}.

\tocless{\subsubsection{Watts--Strogatz networks (\figref{fig:thresholding_comparison} and \ref{fig:swp})}}

\label{methods:WS_network_details}

The original Watts--Strogatz network \cite{Watts1998} consists of a regular lattice basis (characterized by a coordination number $k$) that is then modified, rewiring each edge with a probability $p$.
For directed networks, we used a generalization of that method implemented in NNGT which is strictly equivalent except for the fact that edges are now directed:
\begin{enumerate}
	\item start from a directed regular lattice with coordination number $k$ and reciprocity $r$ (taken as 1 in this paper),
	\item rewire each edge with probability $p$.
\end{enumerate}

The original lattice $L(N, k, r)$ has $\frac{1}{2}Nk (1 + r)$ edges, leading to the limit cases
\begin{itemize}
	\item $\frac{1}{2}Nk$ edges if $r = 0$, like the undirected lattice,
	\item $Nk$ edges if $r = 1$, with all connections being reciprocal.
\end{itemize}

On \figref{fig:thresholding_comparison}D and \ref{fig:swp}A--F (ground-truth graphs), the parameters used are $N = 1000$, $k = 20$, and $r = 0.03$.
The weights associated to the edges are drawn from a shifted exponential distribution $SE(w_{\text{min}}, \lambda) = w_{\text{min}} + Exp(\lambda)$ with $w_{min} = 0.1$ and $\lambda = 1.5$.

The function used in the code is: \\ \href{https://nngt.readthedocs.io/en/stable/modules/generation.html#nngt.generation.watts_strogatz}{\texttt{nngt.generation.watts\_strogatz}}.

\tocless{\subsubsection{Erd\H{o}s--Rényi and scale-free networks (\figref{fig:thresholding_comparison})}}

On \figref{fig:thresholding_comparison} E, the Erd\H{o}s--Rényi graph is $ER(N=1000, E=10000)$.

The function used in the code is: \\ \href{https://nngt.readthedocs.io/en/stable/modules/generation.html#nngt.generation.erdos_renyi}{\texttt{nngt.generation.erdos\_renyi}}.

On \figref{fig:thresholding_comparison} F, the scale-free graph is generated from a Price network \cite{Price1976}.
The graph contains 1000 nodes that are added one after the other, each making at most $m = 10$ out-going edges (for the first 10 nodes, the $i$th only makes $i - 1$ out-going edges).
These edges target already existing nodes with a probability $p \propto d_{\text{in}}^\gamma + c$, which leads to an in-degree distribution for $\gamma = 1$:

\begin{equation}
	P_{d_{\text{in}}} \sim d_{\text{in}}^{-(2 + c/m)}.
\end{equation}

The function used in the code is: \\  \href{https://nngt.readthedocs.io/en/stable/modules/generation.html#nngt.generation.price_scale_free}{\texttt{nngt.generation.price\_scale\_free}}.

\tocless{\subsubsection{Graph shuffling (\figref{fig:brain-motifs})}}

\label{subsubsec:shuffling}

On \figref{fig:brain-motifs} A--D, the influence of the weights on the clustering motifs is assessed by comparing the original connectomes to null-models with the same binary connectivity but shuffled weights.
In practice, from the initial weighted graph $G(V, E, W)$, one creates $G'(V, E, W')$ such that the new weights $W' = \{w'_{ij}\}$ are a random permutation of the original weights $W = \{w_{ij}\}$.

For \figref{fig:brain-motifs} E, the combined effect of weights and binary structure is assessed by comparing the original connectome to an Erd\H{o}s-Rényi graph with the same number of nodes, edges, and the same set of weights.
In practice, from the initial weighted graph $G(V, E, W)$, one creates $ER(V, E', W)$ such that $\vert E \vert = \vert E' \vert$.

\subsection{Small-world propensity}

\label{methods-swp}

As in \cite{Muldoon2016}, the Small-World Propensity (SWP) of a graph $G$ is defined as:
\begin{equation}
	\text{SWP}(G) = 1 - \sqrt{\frac{\Delta_C^2 + \Delta_L^2}{2}}
\end{equation}
with the deviations
\begin{equation}
	\Delta_C = \frac{\langle C^{tot}_l \rangle - \left\langle C^{tot}_G \right\rangle}{\langle C^{tot}_l \rangle - \langle C^{tot}_r\rangle} \quad\text{and}\quad \Delta_L = \frac{\langle L_G \rangle - \langle L_r \rangle}{\langle L_l\rangle - \langle L_r\rangle}
\end{equation}
where $\langle C^{tot}_{l/r} \rangle$ is the average total clustering coefficient of latticized/randomized version of $G$ and $\langle L_{l/r} \rangle$ is the associated average path length.
The deviations are restricted to the $[0, 1]$ interval and clipped if they go below zero or above one.

The main difference, besides the use of directed clustering coefficients and paths, lays in the way ``latticization'' is implemented.
Indeed, as the lattice is used as the reference for the maximum local clustering, we used the following algorithm: from the graph $G$ with $N$ nodes and $E$ edges,
\begin{enumerate}
	\item a fully reciprocal circular of coordination number $k = \lfloor E / 2N\rfloor$ is made
	\item we define the distance between two nodes via the positive modulo $d_{ij} = \vert i - j \vert  \mod N$,
	\item starting from node 0,  the largest weight is attributed to its connection to node 1, then the second largest weight to the reciprocal connection, (1, 0), this is continued with node 1 and connections (1, 2), (2, 1), then with subsequent nodes and connections in a clockwise fashion until all first neighbors are connected
	\item this pattern is continued until all $k$-th nearest neighbors have been connected with decreasing weight values
	\item the remaining $E_r = E - 2N \lfloor E / 2N\rfloor < N$ edges are used to connect the first $\lfloor E_r /2 \rfloor$ nodes to their clockwise $k + 1$-th nearest neighbors, and reciprocally, with decreasing edge strength (note that the last connection will not be reciprocated if $E$ is odd)
\end{enumerate}
This leads to a lattice that maximizes the local clustering as the edges of close neighbors are all reciprocated and associated to large weights.

\section*{Acknowledgments}

Tanguy Fardet received funding from a Humboldt Research Fellowship for Postdoctoral Researchers (Alexander von Humboldt Foundation), then from the European Union’s Horizon Europe research and innovation programme under the Marie Sklodowska-Curie grant agreement No \href{https://cordis.europa.eu/project/id/101063239}{101063239}.
Emmanouil Giannakakis thanks the International Max Planck Research School for Intelligent Systems (IMPRS-IS) and the T\"ubingen AI center for support. Anna Levina is a member of the Machine Learning Cluster of Excellence, EXC number 2064/1 – Project number 39072764. This work was supported by a Sofja Kovalevskaja Award from the Alexander von Humboldt Foundation
We also thank Janne Lappalainen and Zinovia Stefanidi for a valuable discussion about drosophila connectomics.

\newpage
\bibliography{bibliography}

\begin{thebibliography}{50}%
\makeatletter
\providecommand \@ifxundefined [1]{%
 \@ifx{#1\undefined}
}%
\providecommand \@ifnum [1]{%
 \ifnum #1\expandafter \@firstoftwo
 \else \expandafter \@secondoftwo
 \fi
}%
\providecommand \@ifx [1]{%
 \ifx #1\expandafter \@firstoftwo
 \else \expandafter \@secondoftwo
 \fi
}%
\providecommand \natexlab [1]{#1}%
\providecommand \enquote  [1]{``#1''}%
\providecommand \bibnamefont  [1]{#1}%
\providecommand \bibfnamefont [1]{#1}%
\providecommand \citenamefont [1]{#1}%
\providecommand \href@noop [0]{\@secondoftwo}%
\providecommand \href [0]{\begingroup \@sanitize@url \@href}%
\providecommand \@href[1]{\@@startlink{#1}\@@href}%
\providecommand \@@href[1]{\endgroup#1\@@endlink}%
\providecommand \@sanitize@url [0]{\catcode `\\12\catcode `\$12\catcode
  `\&12\catcode `\#12\catcode `\^12\catcode `\_12\catcode `\%12\relax}%
\providecommand \@@startlink[1]{}%
\providecommand \@@endlink[0]{}%
\providecommand \url  [0]{\begingroup\@sanitize@url \@url }%
\providecommand \@url [1]{\endgroup\@href {#1}{\urlprefix }}%
\providecommand \urlprefix  [0]{URL }%
\providecommand \Eprint [0]{\href }%
\providecommand \doibase [0]{https://doi.org/}%
\providecommand \selectlanguage [0]{\@gobble}%
\providecommand \bibinfo  [0]{\@secondoftwo}%
\providecommand \bibfield  [0]{\@secondoftwo}%
\providecommand \translation [1]{[#1]}%
\providecommand \BibitemOpen [0]{}%
\providecommand \bibitemStop [0]{}%
\providecommand \bibitemNoStop [0]{.\EOS\space}%
\providecommand \EOS [0]{\spacefactor3000\relax}%
\providecommand \BibitemShut  [1]{\csname bibitem#1\endcsname}%
\let\auto@bib@innerbib\@empty
\bibitem [{\citenamefont {Wang}\ and\ \citenamefont
  {Kennedy}(2016)}]{Wang.2016}%
  \BibitemOpen
  \bibfield  {author} {\bibinfo {author} {\bibfnamefont {X.-J.}\ \bibnamefont
  {Wang}}\ and\ \bibinfo {author} {\bibfnamefont {H.}~\bibnamefont {Kennedy}},\
  }\bibfield  {title} {\bibinfo {title} {Brain structure and dynamics across
  scales: in search of rules},\ }\href
  {https://doi.org/10.1016/j.conb.2015.12.010} {\bibfield  {journal} {\bibinfo
  {journal} {Current opinion in neurobiology}\ }\textbf {\bibinfo {volume}
  {37}},\ \bibinfo {pages} {92} (\bibinfo {year} {2016})}\BibitemShut {NoStop}%
\bibitem [{\citenamefont {Bonifazi}\ and\ \citenamefont
  {Massobrio}(2019)}]{Bonifazi.2019}%
  \BibitemOpen
  \bibfield  {author} {\bibinfo {author} {\bibfnamefont {P.}~\bibnamefont
  {Bonifazi}}\ and\ \bibinfo {author} {\bibfnamefont {P.}~\bibnamefont
  {Massobrio}},\ }\bibfield  {title} {\bibinfo {title} {Reconstruction of
  functional connectivity from multielectrode recordings and calcium imaging},\
  }\href {https://doi.org/10.1007/978-3-030-11135-9$\backslash$textunderscore}
  {\bibfield  {journal} {\bibinfo  {journal} {Advances in neurobiology}\
  }\textbf {\bibinfo {volume} {22}},\ \bibinfo {pages} {207} (\bibinfo {year}
  {2019})}\BibitemShut {NoStop}%
\bibitem [{\citenamefont {Brookes}\ \emph {et~al.}(2011)\citenamefont
  {Brookes}, \citenamefont {Hale}, \citenamefont {Zumer}, \citenamefont
  {Stevenson}, \citenamefont {Francis}, \citenamefont {Barnes}, \citenamefont
  {Owen}, \citenamefont {Morris},\ and\ \citenamefont
  {Nagarajan}}]{Brookes.2011}%
  \BibitemOpen
  \bibfield  {author} {\bibinfo {author} {\bibfnamefont {M.~J.}\ \bibnamefont
  {Brookes}}, \bibinfo {author} {\bibfnamefont {J.~R.}\ \bibnamefont {Hale}},
  \bibinfo {author} {\bibfnamefont {J.~M.}\ \bibnamefont {Zumer}}, \bibinfo
  {author} {\bibfnamefont {C.~M.}\ \bibnamefont {Stevenson}}, \bibinfo {author}
  {\bibfnamefont {S.~T.}\ \bibnamefont {Francis}}, \bibinfo {author}
  {\bibfnamefont {G.~R.}\ \bibnamefont {Barnes}}, \bibinfo {author}
  {\bibfnamefont {J.~P.}\ \bibnamefont {Owen}}, \bibinfo {author}
  {\bibfnamefont {P.~G.}\ \bibnamefont {Morris}},\ and\ \bibinfo {author}
  {\bibfnamefont {S.~S.}\ \bibnamefont {Nagarajan}},\ }\bibfield  {title}
  {\bibinfo {title} {Measuring functional connectivity using meg: methodology
  and comparison with fcmri},\ }\href
  {https://doi.org/10.1016/j.neuroimage.2011.02.054} {\bibfield  {journal}
  {\bibinfo  {journal} {NeuroImage}\ }\textbf {\bibinfo {volume} {56}},\
  \bibinfo {pages} {1082} (\bibinfo {year} {2011})}\BibitemShut {NoStop}%
\bibitem [{\citenamefont {Cook}\ \emph {et~al.}(2019)\citenamefont {Cook},
  \citenamefont {Jarrell}, \citenamefont {Brittin}, \citenamefont {Wang},
  \citenamefont {Bloniarz}, \citenamefont {Yakovlev}, \citenamefont {Nguyen},
  \citenamefont {Tang}, \citenamefont {Bayer}, \citenamefont {Duerr},
  \citenamefont {B{\"u}low}, \citenamefont {Hobert}, \citenamefont {Hall},\
  and\ \citenamefont {Emmons}}]{Cook.2019}%
  \BibitemOpen
  \bibfield  {author} {\bibinfo {author} {\bibfnamefont {S.~J.}\ \bibnamefont
  {Cook}}, \bibinfo {author} {\bibfnamefont {T.~A.}\ \bibnamefont {Jarrell}},
  \bibinfo {author} {\bibfnamefont {C.~A.}\ \bibnamefont {Brittin}}, \bibinfo
  {author} {\bibfnamefont {Y.}~\bibnamefont {Wang}}, \bibinfo {author}
  {\bibfnamefont {A.~E.}\ \bibnamefont {Bloniarz}}, \bibinfo {author}
  {\bibfnamefont {M.~A.}\ \bibnamefont {Yakovlev}}, \bibinfo {author}
  {\bibfnamefont {K.~C.~Q.}\ \bibnamefont {Nguyen}}, \bibinfo {author}
  {\bibfnamefont {L.~T.-H.}\ \bibnamefont {Tang}}, \bibinfo {author}
  {\bibfnamefont {E.~A.}\ \bibnamefont {Bayer}}, \bibinfo {author}
  {\bibfnamefont {J.~S.}\ \bibnamefont {Duerr}}, \bibinfo {author}
  {\bibfnamefont {H.~E.}\ \bibnamefont {B{\"u}low}}, \bibinfo {author}
  {\bibfnamefont {O.}~\bibnamefont {Hobert}}, \bibinfo {author} {\bibfnamefont
  {D.~H.}\ \bibnamefont {Hall}},\ and\ \bibinfo {author} {\bibfnamefont
  {S.~W.}\ \bibnamefont {Emmons}},\ }\bibfield  {title} {\bibinfo {title}
  {Whole-animal connectomes of both caenorhabditis elegans sexes},\ }\href
  {https://doi.org/10.1038/s41586-019-1352-7} {\bibfield  {journal} {\bibinfo
  {journal} {Nature}\ }\textbf {\bibinfo {volume} {571}},\ \bibinfo {pages}
  {63} (\bibinfo {year} {2019})}\BibitemShut {NoStop}%
\bibitem [{\citenamefont {Scheffer}\ \emph {et~al.}(2020)\citenamefont
  {Scheffer}, \citenamefont {Xu}, \citenamefont {Januszewski}, \citenamefont
  {Lu}, \citenamefont {Takemura}, \citenamefont {Hayworth}, \citenamefont
  {Huang}, \citenamefont {Shinomiya}, \citenamefont {Maitlin-Shepard},
  \citenamefont {Berg}, \citenamefont {Clements}, \citenamefont {Hubbard},
  \citenamefont {Katz}, \citenamefont {Umayam}, \citenamefont {Zhao},
  \citenamefont {Ackerman}, \citenamefont {Blakely}, \citenamefont {Bogovic},
  \citenamefont {Dolafi}, \citenamefont {Kainmueller}, \citenamefont {Kawase},
  \citenamefont {Khairy}, \citenamefont {Leavitt}, \citenamefont {Li},
  \citenamefont {Lindsey}, \citenamefont {Neubarth}, \citenamefont {Olbris},
  \citenamefont {Otsuna}, \citenamefont {Trautman}, \citenamefont {Ito},
  \citenamefont {Bates}, \citenamefont {Goldammer}, \citenamefont {Wolff},
  \citenamefont {Svirskas}, \citenamefont {Schlegel}, \citenamefont {Neace},
  \citenamefont {Knecht}, \citenamefont {Alvarado}, \citenamefont {Bailey},
  \citenamefont {Ballinger}, \citenamefont {Borycz}, \citenamefont {Canino},
  \citenamefont {Cheatham}, \citenamefont {Cook}, \citenamefont {Dreher},
  \citenamefont {Duclos}, \citenamefont {Eubanks}, \citenamefont {Fairbanks},
  \citenamefont {Finley}, \citenamefont {Forknall}, \citenamefont {Francis},
  \citenamefont {Hopkins}, \citenamefont {Joyce}, \citenamefont {Kim},
  \citenamefont {Kirk}, \citenamefont {Kovalyak}, \citenamefont {Lauchie},
  \citenamefont {Lohff}, \citenamefont {Maldonado}, \citenamefont {Manley},
  \citenamefont {McLin}, \citenamefont {Mooney}, \citenamefont {Ndama},
  \citenamefont {Ogundeyi}, \citenamefont {Okeoma}, \citenamefont {Ordish},
  \citenamefont {Padilla}, \citenamefont {Patrick}, \citenamefont {Paterson},
  \citenamefont {Phillips}, \citenamefont {Phillips}, \citenamefont {Rampally},
  \citenamefont {Ribeiro}, \citenamefont {Robertson}, \citenamefont {Rymer},
  \citenamefont {Ryan}, \citenamefont {Sammons}, \citenamefont {Scott},
  \citenamefont {Scott}, \citenamefont {Shinomiya}, \citenamefont {Smith},
  \citenamefont {Smith}, \citenamefont {Smith}, \citenamefont {Sobeski},
  \citenamefont {Suleiman}, \citenamefont {Swift}, \citenamefont {Takemura},
  \citenamefont {Talebi}, \citenamefont {Tarnogorska}, \citenamefont {Tenshaw},
  \citenamefont {Tokhi}, \citenamefont {Walsh}, \citenamefont {Yang},
  \citenamefont {Horne}, \citenamefont {Li}, \citenamefont {Parekh},
  \citenamefont {Rivlin}, \citenamefont {Jayaraman}, \citenamefont {Costa},
  \citenamefont {Jefferis}, \citenamefont {Ito}, \citenamefont {Saalfeld},
  \citenamefont {George}, \citenamefont {Meinertzhagen}, \citenamefont {Rubin},
  \citenamefont {Hess}, \citenamefont {Jain},\ and\ \citenamefont
  {Plaza}}]{Scheffer.2020}%
  \BibitemOpen
  \bibfield  {author} {\bibinfo {author} {\bibfnamefont {L.~K.}\ \bibnamefont
  {Scheffer}}, \bibinfo {author} {\bibfnamefont {C.~S.}\ \bibnamefont {Xu}},
  \bibinfo {author} {\bibfnamefont {M.}~\bibnamefont {Januszewski}}, \bibinfo
  {author} {\bibfnamefont {Z.}~\bibnamefont {Lu}}, \bibinfo {author}
  {\bibfnamefont {S.-Y.}\ \bibnamefont {Takemura}}, \bibinfo {author}
  {\bibfnamefont {K.~J.}\ \bibnamefont {Hayworth}}, \bibinfo {author}
  {\bibfnamefont {G.~B.}\ \bibnamefont {Huang}}, \bibinfo {author}
  {\bibfnamefont {K.}~\bibnamefont {Shinomiya}}, \bibinfo {author}
  {\bibfnamefont {J.}~\bibnamefont {Maitlin-Shepard}}, \bibinfo {author}
  {\bibfnamefont {S.}~\bibnamefont {Berg}}, \bibinfo {author} {\bibfnamefont
  {J.}~\bibnamefont {Clements}}, \bibinfo {author} {\bibfnamefont {P.~M.}\
  \bibnamefont {Hubbard}}, \bibinfo {author} {\bibfnamefont {W.~T.}\
  \bibnamefont {Katz}}, \bibinfo {author} {\bibfnamefont {L.}~\bibnamefont
  {Umayam}}, \bibinfo {author} {\bibfnamefont {T.}~\bibnamefont {Zhao}},
  \bibinfo {author} {\bibfnamefont {D.}~\bibnamefont {Ackerman}}, \bibinfo
  {author} {\bibfnamefont {T.}~\bibnamefont {Blakely}}, \bibinfo {author}
  {\bibfnamefont {J.}~\bibnamefont {Bogovic}}, \bibinfo {author} {\bibfnamefont
  {T.}~\bibnamefont {Dolafi}}, \bibinfo {author} {\bibfnamefont
  {D.}~\bibnamefont {Kainmueller}}, \bibinfo {author} {\bibfnamefont
  {T.}~\bibnamefont {Kawase}}, \bibinfo {author} {\bibfnamefont {K.~A.}\
  \bibnamefont {Khairy}}, \bibinfo {author} {\bibfnamefont {L.}~\bibnamefont
  {Leavitt}}, \bibinfo {author} {\bibfnamefont {P.~H.}\ \bibnamefont {Li}},
  \bibinfo {author} {\bibfnamefont {L.}~\bibnamefont {Lindsey}}, \bibinfo
  {author} {\bibfnamefont {N.}~\bibnamefont {Neubarth}}, \bibinfo {author}
  {\bibfnamefont {D.~J.}\ \bibnamefont {Olbris}}, \bibinfo {author}
  {\bibfnamefont {H.}~\bibnamefont {Otsuna}}, \bibinfo {author} {\bibfnamefont
  {E.~T.}\ \bibnamefont {Trautman}}, \bibinfo {author} {\bibfnamefont
  {M.}~\bibnamefont {Ito}}, \bibinfo {author} {\bibfnamefont {A.~S.}\
  \bibnamefont {Bates}}, \bibinfo {author} {\bibfnamefont {J.}~\bibnamefont
  {Goldammer}}, \bibinfo {author} {\bibfnamefont {T.}~\bibnamefont {Wolff}},
  \bibinfo {author} {\bibfnamefont {R.}~\bibnamefont {Svirskas}}, \bibinfo
  {author} {\bibfnamefont {P.}~\bibnamefont {Schlegel}}, \bibinfo {author}
  {\bibfnamefont {E.}~\bibnamefont {Neace}}, \bibinfo {author} {\bibfnamefont
  {C.~J.}\ \bibnamefont {Knecht}}, \bibinfo {author} {\bibfnamefont {C.~X.}\
  \bibnamefont {Alvarado}}, \bibinfo {author} {\bibfnamefont {D.~A.}\
  \bibnamefont {Bailey}}, \bibinfo {author} {\bibfnamefont {S.}~\bibnamefont
  {Ballinger}}, \bibinfo {author} {\bibfnamefont {J.~A.}\ \bibnamefont
  {Borycz}}, \bibinfo {author} {\bibfnamefont {B.~S.}\ \bibnamefont {Canino}},
  \bibinfo {author} {\bibfnamefont {N.}~\bibnamefont {Cheatham}}, \bibinfo
  {author} {\bibfnamefont {M.}~\bibnamefont {Cook}}, \bibinfo {author}
  {\bibfnamefont {M.}~\bibnamefont {Dreher}}, \bibinfo {author} {\bibfnamefont
  {O.}~\bibnamefont {Duclos}}, \bibinfo {author} {\bibfnamefont
  {B.}~\bibnamefont {Eubanks}}, \bibinfo {author} {\bibfnamefont
  {K.}~\bibnamefont {Fairbanks}}, \bibinfo {author} {\bibfnamefont
  {S.}~\bibnamefont {Finley}}, \bibinfo {author} {\bibfnamefont
  {N.}~\bibnamefont {Forknall}}, \bibinfo {author} {\bibfnamefont
  {A.}~\bibnamefont {Francis}}, \bibinfo {author} {\bibfnamefont {G.~P.}\
  \bibnamefont {Hopkins}}, \bibinfo {author} {\bibfnamefont {E.~M.}\
  \bibnamefont {Joyce}}, \bibinfo {author} {\bibfnamefont {S.}~\bibnamefont
  {Kim}}, \bibinfo {author} {\bibfnamefont {N.~A.}\ \bibnamefont {Kirk}},
  \bibinfo {author} {\bibfnamefont {J.}~\bibnamefont {Kovalyak}}, \bibinfo
  {author} {\bibfnamefont {S.~A.}\ \bibnamefont {Lauchie}}, \bibinfo {author}
  {\bibfnamefont {A.}~\bibnamefont {Lohff}}, \bibinfo {author} {\bibfnamefont
  {C.}~\bibnamefont {Maldonado}}, \bibinfo {author} {\bibfnamefont {E.~A.}\
  \bibnamefont {Manley}}, \bibinfo {author} {\bibfnamefont {S.}~\bibnamefont
  {McLin}}, \bibinfo {author} {\bibfnamefont {C.}~\bibnamefont {Mooney}},
  \bibinfo {author} {\bibfnamefont {M.}~\bibnamefont {Ndama}}, \bibinfo
  {author} {\bibfnamefont {O.}~\bibnamefont {Ogundeyi}}, \bibinfo {author}
  {\bibfnamefont {N.}~\bibnamefont {Okeoma}}, \bibinfo {author} {\bibfnamefont
  {C.}~\bibnamefont {Ordish}}, \bibinfo {author} {\bibfnamefont
  {N.}~\bibnamefont {Padilla}}, \bibinfo {author} {\bibfnamefont {C.~M.}\
  \bibnamefont {Patrick}}, \bibinfo {author} {\bibfnamefont {T.}~\bibnamefont
  {Paterson}}, \bibinfo {author} {\bibfnamefont {E.~E.}\ \bibnamefont
  {Phillips}}, \bibinfo {author} {\bibfnamefont {E.~M.}\ \bibnamefont
  {Phillips}}, \bibinfo {author} {\bibfnamefont {N.}~\bibnamefont {Rampally}},
  \bibinfo {author} {\bibfnamefont {C.}~\bibnamefont {Ribeiro}}, \bibinfo
  {author} {\bibfnamefont {M.~K.}\ \bibnamefont {Robertson}}, \bibinfo {author}
  {\bibfnamefont {J.~T.}\ \bibnamefont {Rymer}}, \bibinfo {author}
  {\bibfnamefont {S.~M.}\ \bibnamefont {Ryan}}, \bibinfo {author}
  {\bibfnamefont {M.}~\bibnamefont {Sammons}}, \bibinfo {author} {\bibfnamefont
  {A.~K.}\ \bibnamefont {Scott}}, \bibinfo {author} {\bibfnamefont {A.~L.}\
  \bibnamefont {Scott}}, \bibinfo {author} {\bibfnamefont {A.}~\bibnamefont
  {Shinomiya}}, \bibinfo {author} {\bibfnamefont {C.}~\bibnamefont {Smith}},
  \bibinfo {author} {\bibfnamefont {K.}~\bibnamefont {Smith}}, \bibinfo
  {author} {\bibfnamefont {N.~L.}\ \bibnamefont {Smith}}, \bibinfo {author}
  {\bibfnamefont {M.~A.}\ \bibnamefont {Sobeski}}, \bibinfo {author}
  {\bibfnamefont {A.}~\bibnamefont {Suleiman}}, \bibinfo {author}
  {\bibfnamefont {J.}~\bibnamefont {Swift}}, \bibinfo {author} {\bibfnamefont
  {S.}~\bibnamefont {Takemura}}, \bibinfo {author} {\bibfnamefont
  {I.}~\bibnamefont {Talebi}}, \bibinfo {author} {\bibfnamefont
  {D.}~\bibnamefont {Tarnogorska}}, \bibinfo {author} {\bibfnamefont
  {E.}~\bibnamefont {Tenshaw}}, \bibinfo {author} {\bibfnamefont
  {T.}~\bibnamefont {Tokhi}}, \bibinfo {author} {\bibfnamefont {J.~J.}\
  \bibnamefont {Walsh}}, \bibinfo {author} {\bibfnamefont {T.}~\bibnamefont
  {Yang}}, \bibinfo {author} {\bibfnamefont {J.~A.}\ \bibnamefont {Horne}},
  \bibinfo {author} {\bibfnamefont {F.}~\bibnamefont {Li}}, \bibinfo {author}
  {\bibfnamefont {R.}~\bibnamefont {Parekh}}, \bibinfo {author} {\bibfnamefont
  {P.~K.}\ \bibnamefont {Rivlin}}, \bibinfo {author} {\bibfnamefont
  {V.}~\bibnamefont {Jayaraman}}, \bibinfo {author} {\bibfnamefont
  {M.}~\bibnamefont {Costa}}, \bibinfo {author} {\bibfnamefont {G.~S.}\
  \bibnamefont {Jefferis}}, \bibinfo {author} {\bibfnamefont {K.}~\bibnamefont
  {Ito}}, \bibinfo {author} {\bibfnamefont {S.}~\bibnamefont {Saalfeld}},
  \bibinfo {author} {\bibfnamefont {R.}~\bibnamefont {George}}, \bibinfo
  {author} {\bibfnamefont {I.~A.}\ \bibnamefont {Meinertzhagen}}, \bibinfo
  {author} {\bibfnamefont {G.~M.}\ \bibnamefont {Rubin}}, \bibinfo {author}
  {\bibfnamefont {H.~F.}\ \bibnamefont {Hess}}, \bibinfo {author}
  {\bibfnamefont {V.}~\bibnamefont {Jain}},\ and\ \bibinfo {author}
  {\bibfnamefont {S.~M.}\ \bibnamefont {Plaza}},\ }\bibfield  {title} {\bibinfo
  {title} {A connectome and analysis of the adult drosophila central brain},\
  }\bibfield  {journal} {\bibinfo  {journal} {eLife}\ }\textbf {\bibinfo
  {volume} {9}},\ \href {https://doi.org/10.7554/eLife.57443}
  {10.7554/eLife.57443} (\bibinfo {year} {2020})\BibitemShut {NoStop}%
\bibitem [{\citenamefont {Winding}\ \emph {et~al.}(2023)\citenamefont
  {Winding}, \citenamefont {Pedigo}, \citenamefont {Barnes}, \citenamefont
  {Patsolic}, \citenamefont {Park}, \citenamefont {Kazimiers}, \citenamefont
  {Fushiki}, \citenamefont {Andrade}, \citenamefont {Khandelwal}, \citenamefont
  {Valdes-Aleman}, \citenamefont {Li}, \citenamefont {Randel}, \citenamefont
  {Barsotti}, \citenamefont {Correia}, \citenamefont {Fetter}, \citenamefont
  {Hartenstein}, \citenamefont {Priebe}, \citenamefont {Vogelstein},
  \citenamefont {Cardona},\ and\ \citenamefont {Zlatic}}]{Winding.2023}%
  \BibitemOpen
  \bibfield  {author} {\bibinfo {author} {\bibfnamefont {M.}~\bibnamefont
  {Winding}}, \bibinfo {author} {\bibfnamefont {B.~D.}\ \bibnamefont {Pedigo}},
  \bibinfo {author} {\bibfnamefont {C.~L.}\ \bibnamefont {Barnes}}, \bibinfo
  {author} {\bibfnamefont {H.~G.}\ \bibnamefont {Patsolic}}, \bibinfo {author}
  {\bibfnamefont {Y.}~\bibnamefont {Park}}, \bibinfo {author} {\bibfnamefont
  {T.}~\bibnamefont {Kazimiers}}, \bibinfo {author} {\bibfnamefont
  {A.}~\bibnamefont {Fushiki}}, \bibinfo {author} {\bibfnamefont {I.~V.}\
  \bibnamefont {Andrade}}, \bibinfo {author} {\bibfnamefont {A.}~\bibnamefont
  {Khandelwal}}, \bibinfo {author} {\bibfnamefont {J.}~\bibnamefont
  {Valdes-Aleman}}, \bibinfo {author} {\bibfnamefont {F.}~\bibnamefont {Li}},
  \bibinfo {author} {\bibfnamefont {N.}~\bibnamefont {Randel}}, \bibinfo
  {author} {\bibfnamefont {E.}~\bibnamefont {Barsotti}}, \bibinfo {author}
  {\bibfnamefont {A.}~\bibnamefont {Correia}}, \bibinfo {author} {\bibfnamefont
  {R.~D.}\ \bibnamefont {Fetter}}, \bibinfo {author} {\bibfnamefont
  {V.}~\bibnamefont {Hartenstein}}, \bibinfo {author} {\bibfnamefont {C.~E.}\
  \bibnamefont {Priebe}}, \bibinfo {author} {\bibfnamefont {J.~T.}\
  \bibnamefont {Vogelstein}}, \bibinfo {author} {\bibfnamefont
  {A.}~\bibnamefont {Cardona}},\ and\ \bibinfo {author} {\bibfnamefont
  {M.}~\bibnamefont {Zlatic}},\ }\bibfield  {title} {\bibinfo {title} {The
  connectome of an insect brain},\ }\href
  {https://doi.org/10.1126/science.add9330} {\bibfield  {journal} {\bibinfo
  {journal} {Science}\ }\textbf {\bibinfo {volume} {379}},\ \bibinfo {pages}
  {eadd9330} (\bibinfo {year} {2023})},\ \Eprint
  {https://arxiv.org/abs/https://www.science.org/doi/pdf/10.1126/science.add9330}
  {https://www.science.org/doi/pdf/10.1126/science.add9330} \BibitemShut
  {NoStop}%
\bibitem [{\citenamefont {Oh}\ \emph {et~al.}(2014)\citenamefont {Oh},
  \citenamefont {Harris}, \citenamefont {Ng}, \citenamefont {Winslow},
  \citenamefont {Cain}, \citenamefont {Mihalas}, \citenamefont {Wang},
  \citenamefont {Lau}, \citenamefont {Kuan}, \citenamefont {Henry},
  \citenamefont {Mortrud}, \citenamefont {Ouellette}, \citenamefont {Nguyen},
  \citenamefont {Sorensen}, \citenamefont {Slaughterbeck}, \citenamefont
  {Wakeman}, \citenamefont {Li}, \citenamefont {Feng}, \citenamefont {Ho},
  \citenamefont {Nicholas}, \citenamefont {Hirokawa}, \citenamefont {Bohn},
  \citenamefont {Joines}, \citenamefont {Peng}, \citenamefont {Hawrylycz},
  \citenamefont {Phillips}, \citenamefont {Hohmann}, \citenamefont {Wohnoutka},
  \citenamefont {Gerfen}, \citenamefont {Koch}, \citenamefont {Bernard},
  \citenamefont {Dang}, \citenamefont {Jones},\ and\ \citenamefont
  {Zeng}}]{Oh.2014}%
  \BibitemOpen
  \bibfield  {author} {\bibinfo {author} {\bibfnamefont {S.~W.}\ \bibnamefont
  {Oh}}, \bibinfo {author} {\bibfnamefont {J.~A.}\ \bibnamefont {Harris}},
  \bibinfo {author} {\bibfnamefont {L.}~\bibnamefont {Ng}}, \bibinfo {author}
  {\bibfnamefont {B.}~\bibnamefont {Winslow}}, \bibinfo {author} {\bibfnamefont
  {N.}~\bibnamefont {Cain}}, \bibinfo {author} {\bibfnamefont {S.}~\bibnamefont
  {Mihalas}}, \bibinfo {author} {\bibfnamefont {Q.}~\bibnamefont {Wang}},
  \bibinfo {author} {\bibfnamefont {C.}~\bibnamefont {Lau}}, \bibinfo {author}
  {\bibfnamefont {L.}~\bibnamefont {Kuan}}, \bibinfo {author} {\bibfnamefont
  {A.~M.}\ \bibnamefont {Henry}}, \bibinfo {author} {\bibfnamefont {M.~T.}\
  \bibnamefont {Mortrud}}, \bibinfo {author} {\bibfnamefont {B.}~\bibnamefont
  {Ouellette}}, \bibinfo {author} {\bibfnamefont {T.~N.}\ \bibnamefont
  {Nguyen}}, \bibinfo {author} {\bibfnamefont {S.~A.}\ \bibnamefont
  {Sorensen}}, \bibinfo {author} {\bibfnamefont {C.~R.}\ \bibnamefont
  {Slaughterbeck}}, \bibinfo {author} {\bibfnamefont {W.}~\bibnamefont
  {Wakeman}}, \bibinfo {author} {\bibfnamefont {Y.}~\bibnamefont {Li}},
  \bibinfo {author} {\bibfnamefont {D.}~\bibnamefont {Feng}}, \bibinfo {author}
  {\bibfnamefont {A.}~\bibnamefont {Ho}}, \bibinfo {author} {\bibfnamefont
  {E.}~\bibnamefont {Nicholas}}, \bibinfo {author} {\bibfnamefont {K.~E.}\
  \bibnamefont {Hirokawa}}, \bibinfo {author} {\bibfnamefont {P.}~\bibnamefont
  {Bohn}}, \bibinfo {author} {\bibfnamefont {K.~M.}\ \bibnamefont {Joines}},
  \bibinfo {author} {\bibfnamefont {H.}~\bibnamefont {Peng}}, \bibinfo {author}
  {\bibfnamefont {M.~J.}\ \bibnamefont {Hawrylycz}}, \bibinfo {author}
  {\bibfnamefont {J.~W.}\ \bibnamefont {Phillips}}, \bibinfo {author}
  {\bibfnamefont {J.~G.}\ \bibnamefont {Hohmann}}, \bibinfo {author}
  {\bibfnamefont {P.}~\bibnamefont {Wohnoutka}}, \bibinfo {author}
  {\bibfnamefont {C.~R.}\ \bibnamefont {Gerfen}}, \bibinfo {author}
  {\bibfnamefont {C.}~\bibnamefont {Koch}}, \bibinfo {author} {\bibfnamefont
  {A.}~\bibnamefont {Bernard}}, \bibinfo {author} {\bibfnamefont
  {C.}~\bibnamefont {Dang}}, \bibinfo {author} {\bibfnamefont {A.~R.}\
  \bibnamefont {Jones}},\ and\ \bibinfo {author} {\bibfnamefont
  {H.}~\bibnamefont {Zeng}},\ }\bibfield  {title} {\bibinfo {title} {A
  mesoscale connectome of the mouse brain},\ }\href
  {https://doi.org/10.1038/nature13186} {\bibfield  {journal} {\bibinfo
  {journal} {Nature}\ }\textbf {\bibinfo {volume} {508}},\ \bibinfo {pages}
  {207} (\bibinfo {year} {2014})}\BibitemShut {NoStop}%
\bibitem [{\citenamefont {Craddock}\ \emph {et~al.}(2013)\citenamefont
  {Craddock}, \citenamefont {Jbabdi}, \citenamefont {Yan}, \citenamefont
  {Vogelstein}, \citenamefont {Castellanos}, \citenamefont {{Di Martino}},
  \citenamefont {Kelly}, \citenamefont {Heberlein}, \citenamefont {Colcombe},\
  and\ \citenamefont {Milham}}]{Craddock.2013}%
  \BibitemOpen
  \bibfield  {author} {\bibinfo {author} {\bibfnamefont {R.~C.}\ \bibnamefont
  {Craddock}}, \bibinfo {author} {\bibfnamefont {S.}~\bibnamefont {Jbabdi}},
  \bibinfo {author} {\bibfnamefont {C.-G.}\ \bibnamefont {Yan}}, \bibinfo
  {author} {\bibfnamefont {J.~T.}\ \bibnamefont {Vogelstein}}, \bibinfo
  {author} {\bibfnamefont {F.~X.}\ \bibnamefont {Castellanos}}, \bibinfo
  {author} {\bibfnamefont {A.}~\bibnamefont {{Di Martino}}}, \bibinfo {author}
  {\bibfnamefont {C.}~\bibnamefont {Kelly}}, \bibinfo {author} {\bibfnamefont
  {K.}~\bibnamefont {Heberlein}}, \bibinfo {author} {\bibfnamefont
  {S.}~\bibnamefont {Colcombe}},\ and\ \bibinfo {author} {\bibfnamefont
  {M.~P.}\ \bibnamefont {Milham}},\ }\bibfield  {title} {\bibinfo {title}
  {Imaging human connectomes at the macroscale},\ }\href
  {https://doi.org/10.1038/nmeth.2482} {\bibfield  {journal} {\bibinfo
  {journal} {Nature methods}\ }\textbf {\bibinfo {volume} {10}},\ \bibinfo
  {pages} {524} (\bibinfo {year} {2013})}\BibitemShut {NoStop}%
\bibitem [{\citenamefont {Bassett}\ and\ \citenamefont
  {Sporns}(2017)}]{Bassett.2017}%
  \BibitemOpen
  \bibfield  {author} {\bibinfo {author} {\bibfnamefont {D.~S.}\ \bibnamefont
  {Bassett}}\ and\ \bibinfo {author} {\bibfnamefont {O.}~\bibnamefont
  {Sporns}},\ }\bibfield  {title} {\bibinfo {title} {Network neuroscience},\
  }\href {https://doi.org/10.1038/nn.4502} {\bibfield  {journal} {\bibinfo
  {journal} {Nature neuroscience}\ }\textbf {\bibinfo {volume} {20}},\ \bibinfo
  {pages} {353} (\bibinfo {year} {2017})}\BibitemShut {NoStop}%
\bibitem [{\citenamefont {Newman}(2001)}]{Newman.2001}%
  \BibitemOpen
  \bibfield  {author} {\bibinfo {author} {\bibfnamefont {M.~E.}\ \bibnamefont
  {Newman}},\ }\bibfield  {title} {\bibinfo {title} {Scientific collaboration
  networks. ii. shortest paths, weighted networks, and centrality},\ }\href
  {https://doi.org/10.1103/PhysRevE.64.016132} {\bibfield  {journal} {\bibinfo
  {journal} {Physical Review E}\ }\textbf {\bibinfo {volume} {64}},\ \bibinfo
  {pages} {016132} (\bibinfo {year} {2001})}\BibitemShut {NoStop}%
\bibitem [{\citenamefont {Muldoon}\ \emph {et~al.}(2016)\citenamefont
  {Muldoon}, \citenamefont {Bridgeford},\ and\ \citenamefont
  {Bassett}}]{Muldoon2016}%
  \BibitemOpen
  \bibfield  {author} {\bibinfo {author} {\bibfnamefont {S.~F.}\ \bibnamefont
  {Muldoon}}, \bibinfo {author} {\bibfnamefont {E.~W.}\ \bibnamefont
  {Bridgeford}},\ and\ \bibinfo {author} {\bibfnamefont {D.~S.}\ \bibnamefont
  {Bassett}},\ }\bibfield  {title} {\bibinfo {title} {Small-world propensity
  and weighted brain networks},\ }\href {https://doi.org/10.1038/srep22057}
  {\bibfield  {journal} {\bibinfo  {journal} {Scientific reports}\ }\textbf
  {\bibinfo {volume} {6}},\ \bibinfo {pages} {22057} (\bibinfo {year}
  {2016})}\BibitemShut {NoStop}%
\bibitem [{\citenamefont {Whitfield-Gabrieli}\ and\ \citenamefont
  {Nieto-Castanon}(2012)}]{Conn}%
  \BibitemOpen
  \bibfield  {author} {\bibinfo {author} {\bibfnamefont {S.}~\bibnamefont
  {Whitfield-Gabrieli}}\ and\ \bibinfo {author} {\bibfnamefont
  {A.}~\bibnamefont {Nieto-Castanon}},\ }\bibfield  {title} {\bibinfo {title}
  {Conn: A functional connectivity toolbox for correlated and anticorrelated
  brain networks},\ }\href {https://doi.org/10.1089/brain.2012.0073} {\bibfield
   {journal} {\bibinfo  {journal} {Brain Connectivity}\ }\textbf {\bibinfo
  {volume} {2}},\ \bibinfo {pages} {125} (\bibinfo {year} {2012})},\ \bibinfo
  {note} {pMID: 22642651},\ \Eprint
  {https://arxiv.org/abs/https://doi.org/10.1089/brain.2012.0073}
  {https://doi.org/10.1089/brain.2012.0073} \BibitemShut {NoStop}%
\bibitem [{\citenamefont {Wang}\ \emph {et~al.}(2015)\citenamefont {Wang},
  \citenamefont {Wang}, \citenamefont {Xia}, \citenamefont {Liao},
  \citenamefont {Evans},\ and\ \citenamefont {He}}]{GRETNA}%
  \BibitemOpen
  \bibfield  {author} {\bibinfo {author} {\bibfnamefont {J.}~\bibnamefont
  {Wang}}, \bibinfo {author} {\bibfnamefont {X.}~\bibnamefont {Wang}}, \bibinfo
  {author} {\bibfnamefont {M.}~\bibnamefont {Xia}}, \bibinfo {author}
  {\bibfnamefont {X.}~\bibnamefont {Liao}}, \bibinfo {author} {\bibfnamefont
  {A.}~\bibnamefont {Evans}},\ and\ \bibinfo {author} {\bibfnamefont
  {Y.}~\bibnamefont {He}},\ }\bibfield  {title} {\bibinfo {title} {Gretna: a
  graph theoretical network analysis toolbox for imaging connectomics},\ }\href
  {https://doi.org/10.3389/fnhum.2015.00386} {\bibfield  {journal} {\bibinfo
  {journal} {Frontiers in Human Neuroscience}\ }\textbf {\bibinfo {volume}
  {9}},\ \bibinfo {pages} {386} (\bibinfo {year} {2015})}\BibitemShut {NoStop}%
\bibitem [{\citenamefont {Waller}\ \emph {et~al.}(2018)\citenamefont {Waller},
  \citenamefont {Brovkin}, \citenamefont {Dorfschmidt}, \citenamefont {Bzdok},
  \citenamefont {Walter},\ and\ \citenamefont {Kruschwitz}}]{GraphVar}%
  \BibitemOpen
  \bibfield  {author} {\bibinfo {author} {\bibfnamefont {L.}~\bibnamefont
  {Waller}}, \bibinfo {author} {\bibfnamefont {A.}~\bibnamefont {Brovkin}},
  \bibinfo {author} {\bibfnamefont {L.}~\bibnamefont {Dorfschmidt}}, \bibinfo
  {author} {\bibfnamefont {D.}~\bibnamefont {Bzdok}}, \bibinfo {author}
  {\bibfnamefont {H.}~\bibnamefont {Walter}},\ and\ \bibinfo {author}
  {\bibfnamefont {J.}~\bibnamefont {Kruschwitz}},\ }\bibfield  {title}
  {\bibinfo {title} {Graphvar 2.0: A user-friendly toolbox for machine learning
  on functional connectivity measures},\ }\href
  {https://doi.org/https://doi.org/10.1016/j.jneumeth.2018.07.001} {\bibfield
  {journal} {\bibinfo  {journal} {Journal of Neuroscience Methods}\ }\textbf
  {\bibinfo {volume} {308}},\ \bibinfo {pages} {21} (\bibinfo {year}
  {2018})}\BibitemShut {NoStop}%
\bibitem [{\citenamefont {Rubinov}\ and\ \citenamefont
  {Sporns}(2010)}]{BrainConnectivityToolbox}%
  \BibitemOpen
  \bibfield  {author} {\bibinfo {author} {\bibfnamefont {M.}~\bibnamefont
  {Rubinov}}\ and\ \bibinfo {author} {\bibfnamefont {O.}~\bibnamefont
  {Sporns}},\ }\bibfield  {title} {\bibinfo {title} {Complex network measures
  of brain connectivity: Uses and interpretations},\ }\href
  {https://doi.org/https://doi.org/10.1016/j.neuroimage.2009.10.003} {\bibfield
   {journal} {\bibinfo  {journal} {NeuroImage}\ }\textbf {\bibinfo {volume}
  {52}},\ \bibinfo {pages} {1059} (\bibinfo {year} {2010})},\ \bibinfo {note}
  {computational Models of the Brain}\BibitemShut {NoStop}%
\bibitem [{\citenamefont {Bielczyk}\ \emph {et~al.}(2018)\citenamefont
  {Bielczyk}, \citenamefont {Walocha}, \citenamefont {Ebel}, \citenamefont
  {Haak}, \citenamefont {Llera}, \citenamefont {Buitelaar}, \citenamefont
  {Glennon},\ and\ \citenamefont {Beckmann}}]{Bielczyk.2018}%
  \BibitemOpen
  \bibfield  {author} {\bibinfo {author} {\bibfnamefont {N.~Z.}\ \bibnamefont
  {Bielczyk}}, \bibinfo {author} {\bibfnamefont {F.}~\bibnamefont {Walocha}},
  \bibinfo {author} {\bibfnamefont {P.~W.}\ \bibnamefont {Ebel}}, \bibinfo
  {author} {\bibfnamefont {K.~V.}\ \bibnamefont {Haak}}, \bibinfo {author}
  {\bibfnamefont {A.}~\bibnamefont {Llera}}, \bibinfo {author} {\bibfnamefont
  {J.~K.}\ \bibnamefont {Buitelaar}}, \bibinfo {author} {\bibfnamefont {J.~C.}\
  \bibnamefont {Glennon}},\ and\ \bibinfo {author} {\bibfnamefont {C.~F.}\
  \bibnamefont {Beckmann}},\ }\bibfield  {title} {\bibinfo {title}
  {Thresholding functional connectomes by means of mixture modeling},\ }\href
  {https://doi.org/10.1016/j.neuroimage.2018.01.003} {\bibfield  {journal}
  {\bibinfo  {journal} {NeuroImage}\ }\textbf {\bibinfo {volume} {171}},\
  \bibinfo {pages} {402} (\bibinfo {year} {2018})}\BibitemShut {NoStop}%
\bibitem [{\citenamefont {Stam}\ \emph {et~al.}(2014)\citenamefont {Stam},
  \citenamefont {Tewarie}, \citenamefont {{van Dellen}}, \citenamefont {{van
  Straaten}}, \citenamefont {Hillebrand},\ and\ \citenamefont {{van
  Mieghem}}}]{Stam.2014}%
  \BibitemOpen
  \bibfield  {author} {\bibinfo {author} {\bibfnamefont {C.~J.}\ \bibnamefont
  {Stam}}, \bibinfo {author} {\bibfnamefont {P.}~\bibnamefont {Tewarie}},
  \bibinfo {author} {\bibfnamefont {E.}~\bibnamefont {{van Dellen}}}, \bibinfo
  {author} {\bibfnamefont {E.~C.~W.}\ \bibnamefont {{van Straaten}}}, \bibinfo
  {author} {\bibfnamefont {A.}~\bibnamefont {Hillebrand}},\ and\ \bibinfo
  {author} {\bibfnamefont {P.}~\bibnamefont {{van Mieghem}}},\ }\bibfield
  {title} {\bibinfo {title} {The trees and the forest: Characterization of
  complex brain networks with minimum spanning trees},\ }\href
  {https://doi.org/10.1016/j.ijpsycho.2014.04.001} {\bibfield  {journal}
  {\bibinfo  {journal} {International journal of psychophysiology : official
  journal of the International Organization of Psychophysiology}\ }\textbf
  {\bibinfo {volume} {92}},\ \bibinfo {pages} {129} (\bibinfo {year}
  {2014})}\BibitemShut {NoStop}%
\bibitem [{\citenamefont {Dimitriadis}\ \emph {et~al.}(2017)\citenamefont
  {Dimitriadis}, \citenamefont {Salis}, \citenamefont {Tarnanas},\ and\
  \citenamefont {Linden}}]{Dimitriadis.2017}%
  \BibitemOpen
  \bibfield  {author} {\bibinfo {author} {\bibfnamefont {S.~I.}\ \bibnamefont
  {Dimitriadis}}, \bibinfo {author} {\bibfnamefont {C.}~\bibnamefont {Salis}},
  \bibinfo {author} {\bibfnamefont {I.}~\bibnamefont {Tarnanas}},\ and\
  \bibinfo {author} {\bibfnamefont {D.~E.}\ \bibnamefont {Linden}},\ }\bibfield
   {title} {\bibinfo {title} {Topological filtering of dynamic functional brain
  networks unfolds informative chronnectomics: A novel data-driven thresholding
  scheme based on orthogonal minimal spanning trees (omsts)},\ }\href
  {https://doi.org/10.3389/fninf.2017.00028} {\bibfield  {journal} {\bibinfo
  {journal} {Frontiers in neuroinformatics}\ }\textbf {\bibinfo {volume}
  {11}},\ \bibinfo {pages} {28} (\bibinfo {year} {2017})}\BibitemShut {NoStop}%
\bibitem [{\citenamefont {Kale}\ \emph {et~al.}(2018)\citenamefont {Kale},
  \citenamefont {Zalesky},\ and\ \citenamefont {Gollo}}]{Kale2018}%
  \BibitemOpen
  \bibfield  {author} {\bibinfo {author} {\bibfnamefont {P.}~\bibnamefont
  {Kale}}, \bibinfo {author} {\bibfnamefont {A.}~\bibnamefont {Zalesky}},\ and\
  \bibinfo {author} {\bibfnamefont {L.~L.}\ \bibnamefont {Gollo}},\ }\bibfield
  {title} {\bibinfo {title} {Estimating the impact of structural
  directionality: {{How}} reliable are undirected connectomes?},\ }\href
  {https://doi.org/10.1162/netn_a_00040} {\bibfield  {journal} {\bibinfo
  {journal} {Network Neuroscience}\ }\textbf {\bibinfo {volume} {2}},\ \bibinfo
  {pages} {259} (\bibinfo {year} {2018})}\BibitemShut {NoStop}%
\bibitem [{\citenamefont {Fardet}\ and\ \citenamefont
  {Levina}(2021)}]{fardetWeightedDirectedClustering2021}%
  \BibitemOpen
  \bibfield  {author} {\bibinfo {author} {\bibfnamefont {T.}~\bibnamefont
  {Fardet}}\ and\ \bibinfo {author} {\bibfnamefont {A.}~\bibnamefont
  {Levina}},\ }\bibfield  {title} {\bibinfo {title} {Weighted directed
  clustering: {{Interpretations}} and requirements for heterogeneous, inferred,
  and measured networks},\ }\href
  {https://doi.org/10.1103/PhysRevResearch.3.043124} {\bibfield  {journal}
  {\bibinfo  {journal} {Physical Review Research}\ }\textbf {\bibinfo {volume}
  {3}},\ \bibinfo {pages} {043124} (\bibinfo {year} {2021})}\BibitemShut
  {NoStop}%
\bibitem [{\citenamefont {Barrat}\ \emph {et~al.}(2004)\citenamefont {Barrat},
  \citenamefont {Barth{\'e}lemy}, \citenamefont {Pastor-Satorras},\ and\
  \citenamefont {Vespignani}}]{Barrat.2004}%
  \BibitemOpen
  \bibfield  {author} {\bibinfo {author} {\bibfnamefont {A.}~\bibnamefont
  {Barrat}}, \bibinfo {author} {\bibfnamefont {M.}~\bibnamefont
  {Barth{\'e}lemy}}, \bibinfo {author} {\bibfnamefont {R.}~\bibnamefont
  {Pastor-Satorras}},\ and\ \bibinfo {author} {\bibfnamefont {A.}~\bibnamefont
  {Vespignani}},\ }\bibfield  {title} {\bibinfo {title} {The architecture of
  complex weighted networks},\ }\href {https://doi.org/10.1073/pnas.0400087101}
  {\bibfield  {journal} {\bibinfo  {journal} {Proceedings of the National
  Academy of Sciences of the United States of America}\ }\textbf {\bibinfo
  {volume} {101}},\ \bibinfo {pages} {3747} (\bibinfo {year}
  {2004})}\BibitemShut {NoStop}%
\bibitem [{\citenamefont {Onnela}\ \emph {et~al.}(2005)\citenamefont {Onnela},
  \citenamefont {Saram{\"a}ki}, \citenamefont {Kert{\'e}sz},\ and\
  \citenamefont {Kaski}}]{Onnela.2005}%
  \BibitemOpen
  \bibfield  {author} {\bibinfo {author} {\bibfnamefont {J.-P.}\ \bibnamefont
  {Onnela}}, \bibinfo {author} {\bibfnamefont {J.}~\bibnamefont
  {Saram{\"a}ki}}, \bibinfo {author} {\bibfnamefont {J.}~\bibnamefont
  {Kert{\'e}sz}},\ and\ \bibinfo {author} {\bibfnamefont {K.}~\bibnamefont
  {Kaski}},\ }\bibfield  {title} {\bibinfo {title} {Intensity and coherence of
  motifs in weighted complex networks},\ }\href
  {https://doi.org/10.1103/PhysRevE.71.065103} {\bibfield  {journal} {\bibinfo
  {journal} {Physical Review E}\ }\textbf {\bibinfo {volume} {71}},\ \bibinfo
  {pages} {065103} (\bibinfo {year} {2005})}\BibitemShut {NoStop}%
\bibitem [{\citenamefont {Clemente}\ and\ \citenamefont
  {Grassi}(2018)}]{Clemente.2018}%
  \BibitemOpen
  \bibfield  {author} {\bibinfo {author} {\bibfnamefont {G.~P.}\ \bibnamefont
  {Clemente}}\ and\ \bibinfo {author} {\bibfnamefont {R.}~\bibnamefont
  {Grassi}},\ }\bibfield  {title} {\bibinfo {title} {Directed clustering in
  weighted networks: A new perspective},\ }\href
  {https://doi.org/10.1016/j.chaos.2017.12.007} {\bibfield  {journal} {\bibinfo
   {journal} {Chaos, Solitons {\&} Fractals}\ }\textbf {\bibinfo {volume}
  {107}},\ \bibinfo {pages} {26} (\bibinfo {year} {2018})}\BibitemShut
  {NoStop}%
\bibitem [{\citenamefont {Fagiolo}(2007)}]{Fagiolo.2007}%
  \BibitemOpen
  \bibfield  {author} {\bibinfo {author} {\bibfnamefont {G.}~\bibnamefont
  {Fagiolo}},\ }\bibfield  {title} {\bibinfo {title} {Clustering in complex
  directed networks},\ }\href {https://doi.org/10.1103/PhysRevE.76.026107}
  {\bibfield  {journal} {\bibinfo  {journal} {Physical Review E}\ }\textbf
  {\bibinfo {volume} {76}},\ \bibinfo {pages} {026107} (\bibinfo {year}
  {2007})}\BibitemShut {NoStop}%
\bibitem [{\citenamefont {Zhang}\ and\ \citenamefont
  {Horvath}(2005)}]{Zhang.2005}%
  \BibitemOpen
  \bibfield  {author} {\bibinfo {author} {\bibfnamefont {B.}~\bibnamefont
  {Zhang}}\ and\ \bibinfo {author} {\bibfnamefont {S.}~\bibnamefont
  {Horvath}},\ }\bibfield  {title} {\bibinfo {title} {A general framework for
  weighted gene co-expression network analysis},\ }\href
  {https://doi.org/10.2202/1544-6115.1128} {\bibfield  {journal} {\bibinfo
  {journal} {Statistical applications in genetics and molecular biology}\
  }\textbf {\bibinfo {volume} {4}},\ \bibinfo {pages} {Article17} (\bibinfo
  {year} {2005})}\BibitemShut {NoStop}%
\bibitem [{\citenamefont {Thomas}\ \emph {et~al.}(2014)\citenamefont {Thomas},
  \citenamefont {Frank}, \citenamefont {Irfanoglu}, \citenamefont {Modi},
  \citenamefont {Saleem}, \citenamefont {Leopold},\ and\ \citenamefont
  {Pierpaoli}}]{thomas2014anatomical}%
  \BibitemOpen
  \bibfield  {author} {\bibinfo {author} {\bibfnamefont {C.}~\bibnamefont
  {Thomas}}, \bibinfo {author} {\bibfnamefont {Q.~Y.}\ \bibnamefont {Frank}},
  \bibinfo {author} {\bibfnamefont {M.~O.}\ \bibnamefont {Irfanoglu}}, \bibinfo
  {author} {\bibfnamefont {P.}~\bibnamefont {Modi}}, \bibinfo {author}
  {\bibfnamefont {K.~S.}\ \bibnamefont {Saleem}}, \bibinfo {author}
  {\bibfnamefont {D.~A.}\ \bibnamefont {Leopold}},\ and\ \bibinfo {author}
  {\bibfnamefont {C.}~\bibnamefont {Pierpaoli}},\ }\bibfield  {title} {\bibinfo
  {title} {Anatomical accuracy of brain connections derived from diffusion mri
  tractography is inherently limited},\ }\href@noop {} {\bibfield  {journal}
  {\bibinfo  {journal} {Proceedings of the National Academy of Sciences}\
  }\textbf {\bibinfo {volume} {111}},\ \bibinfo {pages} {16574} (\bibinfo
  {year} {2014})}\BibitemShut {NoStop}%
\bibitem [{\citenamefont {Zalesky}\ \emph {et~al.}(2016)\citenamefont
  {Zalesky}, \citenamefont {Fornito}, \citenamefont {Cocchi}, \citenamefont
  {Gollo}, \citenamefont {{van den Heuvel}},\ and\ \citenamefont
  {Breakspear}}]{Zalesky.2016}%
  \BibitemOpen
  \bibfield  {author} {\bibinfo {author} {\bibfnamefont {A.}~\bibnamefont
  {Zalesky}}, \bibinfo {author} {\bibfnamefont {A.}~\bibnamefont {Fornito}},
  \bibinfo {author} {\bibfnamefont {L.}~\bibnamefont {Cocchi}}, \bibinfo
  {author} {\bibfnamefont {L.~L.}\ \bibnamefont {Gollo}}, \bibinfo {author}
  {\bibfnamefont {M.~P.}\ \bibnamefont {{van den Heuvel}}},\ and\ \bibinfo
  {author} {\bibfnamefont {M.}~\bibnamefont {Breakspear}},\ }\bibfield  {title}
  {\bibinfo {title} {Connectome sensitivity or specificity: which is more
  important?},\ }\href {https://doi.org/10.1016/j.neuroimage.2016.06.035}
  {\bibfield  {journal} {\bibinfo  {journal} {NeuroImage}\ }\textbf {\bibinfo
  {volume} {142}},\ \bibinfo {pages} {407} (\bibinfo {year}
  {2016})}\BibitemShut {NoStop}%
\bibitem [{\citenamefont {Buzs{\'a}ki}\ and\ \citenamefont
  {Mizuseki}(2014)}]{buzsaki2014log}%
  \BibitemOpen
  \bibfield  {author} {\bibinfo {author} {\bibfnamefont {G.}~\bibnamefont
  {Buzs{\'a}ki}}\ and\ \bibinfo {author} {\bibfnamefont {K.}~\bibnamefont
  {Mizuseki}},\ }\bibfield  {title} {\bibinfo {title} {The log-dynamic brain:
  how skewed distributions affect network operations},\ }\href@noop {}
  {\bibfield  {journal} {\bibinfo  {journal} {Nature Reviews Neuroscience}\
  }\textbf {\bibinfo {volume} {15}},\ \bibinfo {pages} {264} (\bibinfo {year}
  {2014})}\BibitemShut {NoStop}%
\bibitem [{\citenamefont {Loewenstein}\ \emph {et~al.}(2011)\citenamefont
  {Loewenstein}, \citenamefont {Kuras},\ and\ \citenamefont
  {Rumpel}}]{loewenstein2011multiplicative}%
  \BibitemOpen
  \bibfield  {author} {\bibinfo {author} {\bibfnamefont {Y.}~\bibnamefont
  {Loewenstein}}, \bibinfo {author} {\bibfnamefont {A.}~\bibnamefont {Kuras}},\
  and\ \bibinfo {author} {\bibfnamefont {S.}~\bibnamefont {Rumpel}},\
  }\bibfield  {title} {\bibinfo {title} {Multiplicative dynamics underlie the
  emergence of the log-normal distribution of spine sizes in the neocortex in
  vivo},\ }\href@noop {} {\bibfield  {journal} {\bibinfo  {journal} {Journal of
  Neuroscience}\ }\textbf {\bibinfo {volume} {31}},\ \bibinfo {pages} {9481}
  (\bibinfo {year} {2011})}\BibitemShut {NoStop}%
\bibitem [{\citenamefont {Ryan}\ \emph {et~al.}(2016)\citenamefont {Ryan},
  \citenamefont {Lu},\ and\ \citenamefont {Meinertzhagen}}]{Ryan2016}%
  \BibitemOpen
  \bibfield  {author} {\bibinfo {author} {\bibfnamefont {K.}~\bibnamefont
  {Ryan}}, \bibinfo {author} {\bibfnamefont {Z.}~\bibnamefont {Lu}},\ and\
  \bibinfo {author} {\bibfnamefont {I.~A.}\ \bibnamefont {Meinertzhagen}},\
  }\bibfield  {title} {\bibinfo {title} {The {{CNS}} connectome of a tadpole
  larva of {{Ciona}} intestinalis ({{L}}.) highlights sidedness in the brain of
  a chordate sibling},\ }\href {https://doi.org/10.7554/eLife.16962} {\bibfield
   {journal} {\bibinfo  {journal} {eLife}\ }\textbf {\bibinfo {volume} {5}},\
  \bibinfo {pages} {e16962} (\bibinfo {year} {2016})}\BibitemShut {NoStop}%
\bibitem [{\citenamefont {Gruters}\ and\ \citenamefont
  {Groh}(2012)}]{grutersSoundsMultisensoryOther2012}%
  \BibitemOpen
  \bibfield  {author} {\bibinfo {author} {\bibfnamefont {K.}~\bibnamefont
  {Gruters}}\ and\ \bibinfo {author} {\bibfnamefont {J.}~\bibnamefont {Groh}},\
  }\bibfield  {title} {\bibinfo {title} {Sounds and beyond: Multisensory and
  other non-auditory signals in the inferior colliculus},\ }\href
  {https://doi.org/10.3389/fncir.2012.00096} {\bibfield  {journal} {\bibinfo
  {journal} {Frontiers in Neural Circuits}\ }\textbf {\bibinfo {volume} {6}},\
  \bibinfo {pages} {96} (\bibinfo {year} {2012})}\BibitemShut {NoStop}%
\bibitem [{\citenamefont {Ito}\ and\ \citenamefont
  {Feldheim}(2018)}]{itoMouseSuperiorColliculus2018}%
  \BibitemOpen
  \bibfield  {author} {\bibinfo {author} {\bibfnamefont {S.}~\bibnamefont
  {Ito}}\ and\ \bibinfo {author} {\bibfnamefont {D.~A.}\ \bibnamefont
  {Feldheim}},\ }\bibfield  {title} {\bibinfo {title} {The {{Mouse Superior
  Colliculus}}: {{An Emerging Model}} for {{Studying Circuit Formation}} and
  {{Function}}},\ }\href {https://doi.org/10.3389/fncir.2018.00010} {\bibfield
  {journal} {\bibinfo  {journal} {Frontiers in Neural Circuits}\ }\textbf
  {\bibinfo {volume} {12}},\ \bibinfo {pages} {10} (\bibinfo {year}
  {2018})}\BibitemShut {NoStop}%
\bibitem [{\citenamefont {Papo}\ \emph {et~al.}(2016)\citenamefont {Papo},
  \citenamefont {Zanin}, \citenamefont {Martínez},\ and\ \citenamefont
  {Buldú}}]{papo_beware_2016}%
  \BibitemOpen
  \bibfield  {author} {\bibinfo {author} {\bibfnamefont {D.}~\bibnamefont
  {Papo}}, \bibinfo {author} {\bibfnamefont {M.}~\bibnamefont {Zanin}},
  \bibinfo {author} {\bibfnamefont {J.~H.}\ \bibnamefont {Martínez}},\ and\
  \bibinfo {author} {\bibfnamefont {J.~M.}\ \bibnamefont {Buldú}},\ }\bibfield
   {title} {\bibinfo {title} {Beware of the {Small}-{World}
  {Neuroscientist}!},\ }\bibfield  {journal} {\bibinfo  {journal} {Frontiers in
  Human Neuroscience}\ }\textbf {\bibinfo {volume} {10}},\ \href
  {https://doi.org/10.3389/fnhum.2016.00096} {10.3389/fnhum.2016.00096}
  (\bibinfo {year} {2016}),\ \bibinfo {note} {publisher: Frontiers}\BibitemShut
  {NoStop}%
\bibitem [{\citenamefont {Avena-Koenigsberger}\ \emph
  {et~al.}(2019)\citenamefont {Avena-Koenigsberger}, \citenamefont {Yan},
  \citenamefont {Kolchinsky}, \citenamefont {Heuvel}, \citenamefont {Hagmann},\
  and\ \citenamefont {Sporns}}]{avena-koenigsberger_spectrum_2019}%
  \BibitemOpen
  \bibfield  {author} {\bibinfo {author} {\bibfnamefont {A.}~\bibnamefont
  {Avena-Koenigsberger}}, \bibinfo {author} {\bibfnamefont {X.}~\bibnamefont
  {Yan}}, \bibinfo {author} {\bibfnamefont {A.}~\bibnamefont {Kolchinsky}},
  \bibinfo {author} {\bibfnamefont {M.~P. v.~d.}\ \bibnamefont {Heuvel}},
  \bibinfo {author} {\bibfnamefont {P.}~\bibnamefont {Hagmann}},\ and\ \bibinfo
  {author} {\bibfnamefont {O.}~\bibnamefont {Sporns}},\ }\bibfield  {title}
  {\bibinfo {title} {A spectrum of routing strategies for brain networks},\
  }\href {https://doi.org/10/gg7jmq} {\bibfield  {journal} {\bibinfo  {journal}
  {PLOS Computational Biology}\ }\textbf {\bibinfo {volume} {15}},\ \bibinfo
  {pages} {e1006833} (\bibinfo {year} {2019})},\ \bibinfo {note} {publisher:
  Public Library of Science}\BibitemShut {NoStop}%
\bibitem [{\citenamefont {Fardet}(2024)}]{NNGT}%
  \BibitemOpen
  \bibfield  {author} {\bibinfo {author} {\bibfnamefont {T.}~\bibnamefont
  {Fardet}},\ }\href {https://doi.org/10.5281/zenodo.10577028} {\bibinfo
  {title} {{NNGT 2.7.1: improved configuration, scipy, and networkx support}}}
  (\bibinfo {year} {2024})\BibitemShut {NoStop}%
\bibitem [{Note1()}]{Note1}%
  \BibitemOpen
  \bibinfo {note} {\protect \href
  {https://wormwiring.org/pages/adjacency.html}{https://wormwiring.org/pages/adjacency.html}}\BibitemShut
  {NoStop}%
\bibitem [{Note2()}]{Note2}%
  \BibitemOpen
  \bibinfo {note} {\protect \href
  {https://elifesciences.org/articles/16962/figures\#SD6-data}{https://elifesciences.org/articles/16962/figures\#SD6-data}}\BibitemShut
  {NoStop}%
\bibitem [{Note3()}]{Note3}%
  \BibitemOpen
  \bibinfo {note} {\protect \href
  {https://neuprint.janelia.org/}{https://neuprint.janelia.org/}}\BibitemShut
  {NoStop}%
\bibitem [{Note4()}]{Note4}%
  \BibitemOpen
  \bibinfo {note} {\protect \href
  {https://github.com/connectome-neuprint/neuprint-python}{https://github.com/connectome-neuprint/neuprint-python}}\BibitemShut
  {NoStop}%
\bibitem [{\citenamefont {Hulse}\ \emph {et~al.}(2020)\citenamefont {Hulse},
  \citenamefont {Haberkern}, \citenamefont {Franconville}, \citenamefont
  {Turner-Evans}, \citenamefont {Takemura}, \citenamefont {Wolff},
  \citenamefont {Noorman}, \citenamefont {Dreher}, \citenamefont {Dan},
  \citenamefont {Parekh}, \citenamefont {Hermundstad}, \citenamefont {Rubin},\
  and\ \citenamefont {Jayaraman}}]{Hulse2020}%
  \BibitemOpen
  \bibfield  {author} {\bibinfo {author} {\bibfnamefont {B.~K.}\ \bibnamefont
  {Hulse}}, \bibinfo {author} {\bibfnamefont {H.}~\bibnamefont {Haberkern}},
  \bibinfo {author} {\bibfnamefont {R.}~\bibnamefont {Franconville}}, \bibinfo
  {author} {\bibfnamefont {D.~B.}\ \bibnamefont {Turner-Evans}}, \bibinfo
  {author} {\bibfnamefont {S.}~\bibnamefont {Takemura}}, \bibinfo {author}
  {\bibfnamefont {T.}~\bibnamefont {Wolff}}, \bibinfo {author} {\bibfnamefont
  {M.}~\bibnamefont {Noorman}}, \bibinfo {author} {\bibfnamefont
  {M.}~\bibnamefont {Dreher}}, \bibinfo {author} {\bibfnamefont
  {C.}~\bibnamefont {Dan}}, \bibinfo {author} {\bibfnamefont {R.}~\bibnamefont
  {Parekh}}, \bibinfo {author} {\bibfnamefont {A.~M.}\ \bibnamefont
  {Hermundstad}}, \bibinfo {author} {\bibfnamefont {G.~M.}\ \bibnamefont
  {Rubin}},\ and\ \bibinfo {author} {\bibfnamefont {V.}~\bibnamefont
  {Jayaraman}},\ }\bibfield  {title} {\bibinfo {title} {A connectome of the
  {{Drosophila}} central complex reveals network motifs suitable for flexible
  navigation and context-dependent action selection},\ }\href
  {https://doi.org/10.1101/2020.12.08.413955} {\bibfield  {journal} {\bibinfo
  {journal} {bioRxiv}\ ,\ \bibinfo {pages} {2020.12.08.413955}} (\bibinfo
  {year} {2020})}\BibitemShut {NoStop}%
\bibitem [{Note5()}]{Note5}%
  \BibitemOpen
  \bibinfo {note} {\protect \href
  {https://neurodata.io/project/connectomes/}{https://neurodata.io/project/connectomes/}}\BibitemShut
  {NoStop}%
\bibitem [{\citenamefont
  {Pailthorpe}(2019)}]{pailthorpeNetworkAnalysisMesoscale2019}%
  \BibitemOpen
  \bibfield  {author} {\bibinfo {author} {\bibfnamefont {B.~A.}\ \bibnamefont
  {Pailthorpe}},\ }\bibfield  {title} {\bibinfo {title} {Network analysis of
  mesoscale mouse brain structural connectome yields modular structure that
  aligns with anatomical regions and sensory pathways},\ }\href
  {https://doi.org/10.1101/755041} {\bibfield  {journal} {\bibinfo  {journal}
  {bioRxiv}\ ,\ \bibinfo {pages} {755041}} (\bibinfo {year}
  {2019})}\BibitemShut {NoStop}%
\bibitem [{\citenamefont {Carleton}\ \emph {et~al.}(2010)\citenamefont
  {Carleton}, \citenamefont {Accolla},\ and\ \citenamefont
  {Simon}}]{carletonCodingMammalianGustatory2010}%
  \BibitemOpen
  \bibfield  {author} {\bibinfo {author} {\bibfnamefont {A.}~\bibnamefont
  {Carleton}}, \bibinfo {author} {\bibfnamefont {R.}~\bibnamefont {Accolla}},\
  and\ \bibinfo {author} {\bibfnamefont {S.~A.}\ \bibnamefont {Simon}},\
  }\bibfield  {title} {\bibinfo {title} {Coding in the mammalian gustatory
  system},\ }\href {https://doi.org/10.1016/j.tins.2010.04.002} {\bibfield
  {journal} {\bibinfo  {journal} {Trends in Neurosciences}\ }\textbf {\bibinfo
  {volume} {33}},\ \bibinfo {pages} {326} (\bibinfo {year} {2010})},\ \Eprint
  {https://arxiv.org/abs/20493563} {20493563} \BibitemShut {NoStop}%
\bibitem [{\citenamefont {Di~Bonito}\ and\ \citenamefont
  {Studer}(2017)}]{dibonitoCellularMolecularUnderpinnings2017}%
  \BibitemOpen
  \bibfield  {author} {\bibinfo {author} {\bibfnamefont {M.}~\bibnamefont
  {Di~Bonito}}\ and\ \bibinfo {author} {\bibfnamefont {M.}~\bibnamefont
  {Studer}},\ }\bibfield  {title} {\bibinfo {title} {Cellular and {{Molecular
  Underpinnings}} of {{Neuronal Assembly}} in the {{Central Auditory System}}
  during {{Mouse Development}}},\ }\href
  {https://doi.org/10.3389/fncir.2017.00018} {\bibfield  {journal} {\bibinfo
  {journal} {Frontiers in Neural Circuits}\ }\textbf {\bibinfo {volume} {11}},\
  \bibinfo {pages} {18} (\bibinfo {year} {2017})},\ \Eprint
  {https://arxiv.org/abs/28469562} {28469562} \BibitemShut {NoStop}%
\bibitem [{\citenamefont
  {De~Castro}(2009)}]{decastroWiringOlfactionCellular2009}%
  \BibitemOpen
  \bibfield  {author} {\bibinfo {author} {\bibfnamefont {F.}~\bibnamefont
  {De~Castro}},\ }\bibfield  {title} {\bibinfo {title} {Wiring olfaction: The
  cellular and molecular mechanisms that guide the development of synaptic
  connections from the nose to the cortex},\ }\href
  {https://doi.org/10.3389/neuro.22.004.2009} {\bibfield  {journal} {\bibinfo
  {journal} {Frontiers in Neuroscience}\ }\textbf {\bibinfo {volume} {3}},\
  \bibinfo {pages} {4} (\bibinfo {year} {2009})}\BibitemShut {NoStop}%
\bibitem [{\citenamefont {Mori}\ \emph {et~al.}(2013)\citenamefont {Mori},
  \citenamefont {Manabe}, \citenamefont {Narikiyo},\ and\ \citenamefont
  {Onisawa}}]{moriOlfactoryConsciousnessGamma2013}%
  \BibitemOpen
  \bibfield  {author} {\bibinfo {author} {\bibfnamefont {K.}~\bibnamefont
  {Mori}}, \bibinfo {author} {\bibfnamefont {H.}~\bibnamefont {Manabe}},
  \bibinfo {author} {\bibfnamefont {K.}~\bibnamefont {Narikiyo}},\ and\
  \bibinfo {author} {\bibfnamefont {N.}~\bibnamefont {Onisawa}},\ }\bibfield
  {title} {\bibinfo {title} {Olfactory consciousness and gamma oscillation
  couplings across the olfactory bulb, olfactory cortex, and orbitofrontal
  cortex},\ }\href {https://doi.org/10.3389/fpsyg.2013.00743} {\bibfield
  {journal} {\bibinfo  {journal} {Frontiers in psychology}\ }\textbf {\bibinfo
  {volume} {4}},\ \bibinfo {pages} {743} (\bibinfo {year} {2013})}\BibitemShut
  {NoStop}%
\bibitem [{\citenamefont {Jeong}\ \emph {et~al.}(2016)\citenamefont {Jeong},
  \citenamefont {Kim}, \citenamefont {Kim}, \citenamefont {Ferrante},
  \citenamefont {Mitra}, \citenamefont {Osten},\ and\ \citenamefont
  {Kim}}]{jeongComparativeThreedimensionalConnectome2016}%
  \BibitemOpen
  \bibfield  {author} {\bibinfo {author} {\bibfnamefont {M.}~\bibnamefont
  {Jeong}}, \bibinfo {author} {\bibfnamefont {Y.}~\bibnamefont {Kim}}, \bibinfo
  {author} {\bibfnamefont {J.}~\bibnamefont {Kim}}, \bibinfo {author}
  {\bibfnamefont {D.~D.}\ \bibnamefont {Ferrante}}, \bibinfo {author}
  {\bibfnamefont {P.~P.}\ \bibnamefont {Mitra}}, \bibinfo {author}
  {\bibfnamefont {P.}~\bibnamefont {Osten}},\ and\ \bibinfo {author}
  {\bibfnamefont {D.}~\bibnamefont {Kim}},\ }\bibfield  {title} {\bibinfo
  {title} {Comparative three-dimensional connectome map of motor cortical
  projections in the mouse brain},\ }\href {https://doi.org/10.1038/srep20072}
  {\bibfield  {journal} {\bibinfo  {journal} {Scientific Reports}\ }\textbf
  {\bibinfo {volume} {6}},\ \bibinfo {pages} {20072} (\bibinfo {year}
  {2016})}\BibitemShut {NoStop}%
\bibitem [{\citenamefont {Watts}\ and\ \citenamefont
  {Strogatz}(2006)}]{Watts1998}%
  \BibitemOpen
  \bibfield  {author} {\bibinfo {author} {\bibfnamefont {D.~J.}\ \bibnamefont
  {Watts}}\ and\ \bibinfo {author} {\bibfnamefont {S.~H.}\ \bibnamefont
  {Strogatz}},\ }\bibinfo {title} {Collective dynamics of 'small-world'
  networks},\ in\ \href {https://doi.org/doi:10.1515/9781400841356.301} {\emph
  {\bibinfo {booktitle} {The Structure and Dynamics of Networks}}}\ (\bibinfo
  {publisher} {Princeton University Press},\ \bibinfo {address} {Princeton},\
  \bibinfo {year} {2006})\ pp.\ \bibinfo {pages} {301--303}\BibitemShut
  {NoStop}%
\bibitem [{\citenamefont {Price}(1976)}]{Price1976}%
  \BibitemOpen
  \bibfield  {author} {\bibinfo {author} {\bibfnamefont {D.~D.~S.}\
  \bibnamefont {Price}},\ }\bibfield  {title} {\bibinfo {title} {A general
  theory of bibliometric and other cumulative advantage processes},\ }\href
  {https://doi.org/10.1002/asi.4630270505} {\bibfield  {journal} {\bibinfo
  {journal} {Journal of the American Society for Information Science}\ }\textbf
  {\bibinfo {volume} {27}},\ \bibinfo {pages} {292} (\bibinfo {year}
  {1976})}\BibitemShut {NoStop}%
\bibitem [{\citenamefont {Eschbach}\ \emph {et~al.}(2020)\citenamefont
  {Eschbach}, \citenamefont {Fushiki}, \citenamefont {Winding}, \citenamefont
  {{Schneider-Mizell}}, \citenamefont {Shao}, \citenamefont {Arruda},
  \citenamefont {Eichler}, \citenamefont {{Valdes-Aleman}}, \citenamefont
  {Ohyama}, \citenamefont {Thum}, \citenamefont {Gerber}, \citenamefont
  {Fetter}, \citenamefont {Truman}, \citenamefont {{Litwin-Kumar}},
  \citenamefont {Cardona},\ and\ \citenamefont
  {Zlatic}}]{eschbachRecurrentArchitectureAdaptive2020}%
  \BibitemOpen
  \bibfield  {author} {\bibinfo {author} {\bibfnamefont {C.}~\bibnamefont
  {Eschbach}}, \bibinfo {author} {\bibfnamefont {A.}~\bibnamefont {Fushiki}},
  \bibinfo {author} {\bibfnamefont {M.}~\bibnamefont {Winding}}, \bibinfo
  {author} {\bibfnamefont {C.~M.}\ \bibnamefont {{Schneider-Mizell}}}, \bibinfo
  {author} {\bibfnamefont {M.}~\bibnamefont {Shao}}, \bibinfo {author}
  {\bibfnamefont {R.}~\bibnamefont {Arruda}}, \bibinfo {author} {\bibfnamefont
  {K.}~\bibnamefont {Eichler}}, \bibinfo {author} {\bibfnamefont
  {J.}~\bibnamefont {{Valdes-Aleman}}}, \bibinfo {author} {\bibfnamefont
  {T.}~\bibnamefont {Ohyama}}, \bibinfo {author} {\bibfnamefont {A.~S.}\
  \bibnamefont {Thum}}, \bibinfo {author} {\bibfnamefont {B.}~\bibnamefont
  {Gerber}}, \bibinfo {author} {\bibfnamefont {R.~D.}\ \bibnamefont {Fetter}},
  \bibinfo {author} {\bibfnamefont {J.~W.}\ \bibnamefont {Truman}}, \bibinfo
  {author} {\bibfnamefont {A.}~\bibnamefont {{Litwin-Kumar}}}, \bibinfo
  {author} {\bibfnamefont {A.}~\bibnamefont {Cardona}},\ and\ \bibinfo {author}
  {\bibfnamefont {M.}~\bibnamefont {Zlatic}},\ }\bibfield  {title} {\bibinfo
  {title} {Recurrent architecture for adaptive regulation of learning in the
  insect brain},\ }\href {https://doi.org/10.1038/s41593-020-0607-9} {\bibfield
   {journal} {\bibinfo  {journal} {Nature Neuroscience}\ }\textbf {\bibinfo
  {volume} {23}},\ \bibinfo {pages} {544} (\bibinfo {year} {2020})}\BibitemShut
  {NoStop}%
\end{thebibliography}%

\newpage
\appendix

\section{Hierarchical structures along the mouse cortical pathways}

\begin{figure*}
	\includegraphics[width=\textwidth]{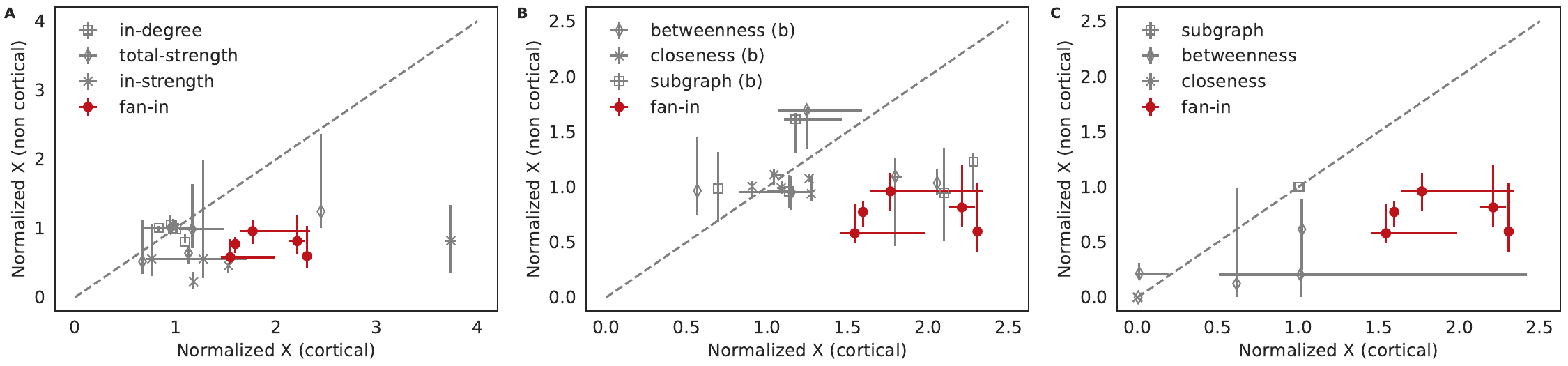}
	\caption{Fully-weighted fan-in clustering is the only graph measure that can consistently identify the cortical areas. This is shown by comparing the normalized values of cortical areas (x-axis) to the normalized values of non-cortical areas (y-axis). Normalization is done by dividing by the average value over all cortical and non-cortical areas. Degree- or strength-based measures do not discriminate the cortical areas (\textbf{A}); neither do standard binary centralities such as betweenness, closeness, and subgraph centrality (\textbf{B}), nor their weighted versions (\textbf{C}).
	}
	\label{fig:supp-mouse-measures-comp}
\end{figure*}

\begin{figure*}
	\includegraphics[width=\textwidth]{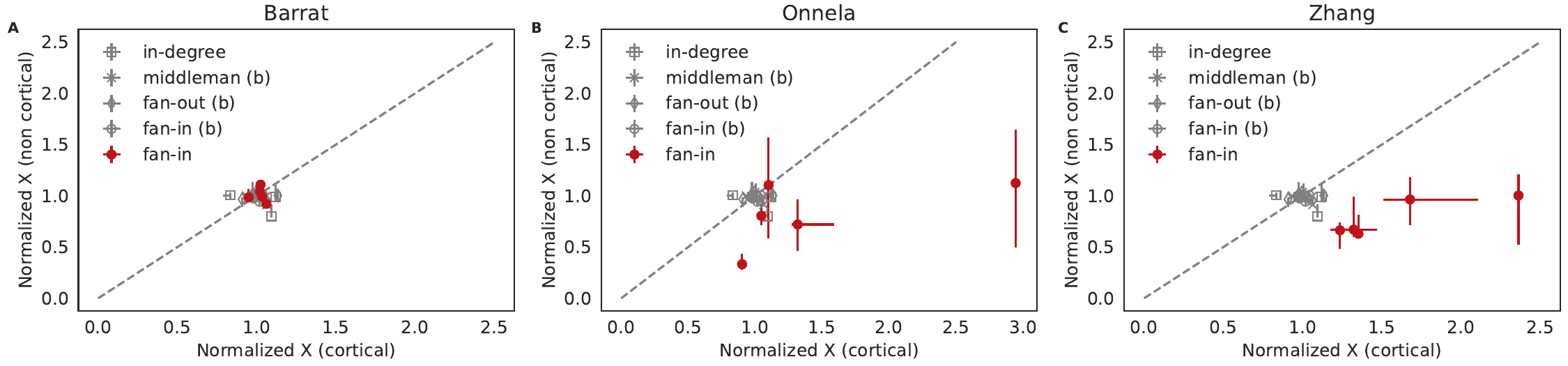}
	\caption{Different methods of estimating the fan-in clustering enable discrimination of isocortical areas from the rest of the pathway areas for the mouse connectome. Hybrid methods (\textbf{A.} Barrat fails to discern any patterns and \textbf{B.} Onella does not allow reliable discrimination) perform much worse compared to fully weighted methods (\textbf{C.} Zhang and continuous \figref{fig:mouse-pathways} D)}
	\label{fig:supp-mouse-fan-in-comp}
\end{figure*}

The hierarchical structures are only visible through fully-weighted fan-in clustering --- Figure~\ref{fig:supp-mouse-fan-in-comp} --- and cannot be identified by other standard graph measurements --- Figure~\ref{fig:supp-mouse-measures-comp}.

\section{Small-world propensity of the \emph{C. elegans} and mouse connectomes}

\begin{figure*}
	\includegraphics[width=\textwidth]{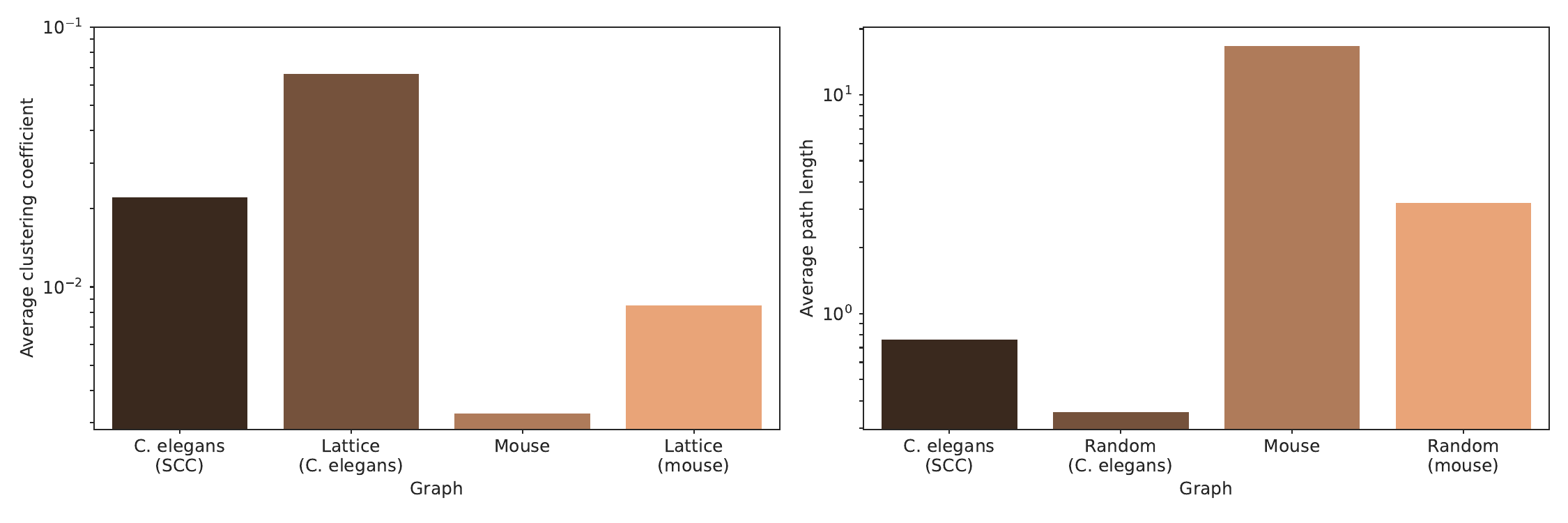}
	\caption{
		Left: compared to a lattice with the same number of nodes, edges, and the same weights, the mouse and C. \textit{elegans} connectomes have much lower average clustering coefficients. Right: similarly, they have a longer average path length than an equivalent random graph. 
  Note the logarithmic scaling of the y-axis. 
	}
	\label{fig:supp_swp}
\end{figure*}

Both \emph{C. elegans} and the mouse mesoscale connectome have simultaneously low weighted directed clustering compared to an equivalent lattice and high average path length compared to an equivalent random graph (cf. shuffling methods \ref{subsubsec:shuffling}), as shown on \supfigref{fig:supp_swp}, which leads to their low SWP on Figure~\ref{fig:swp}. Note the logarithmic scale of the y-axis. 

\section{Clustering patterns correlate with modulatory structures in the drosophila connectome}

\begin{figure*}
	\includegraphics[width=\textwidth]{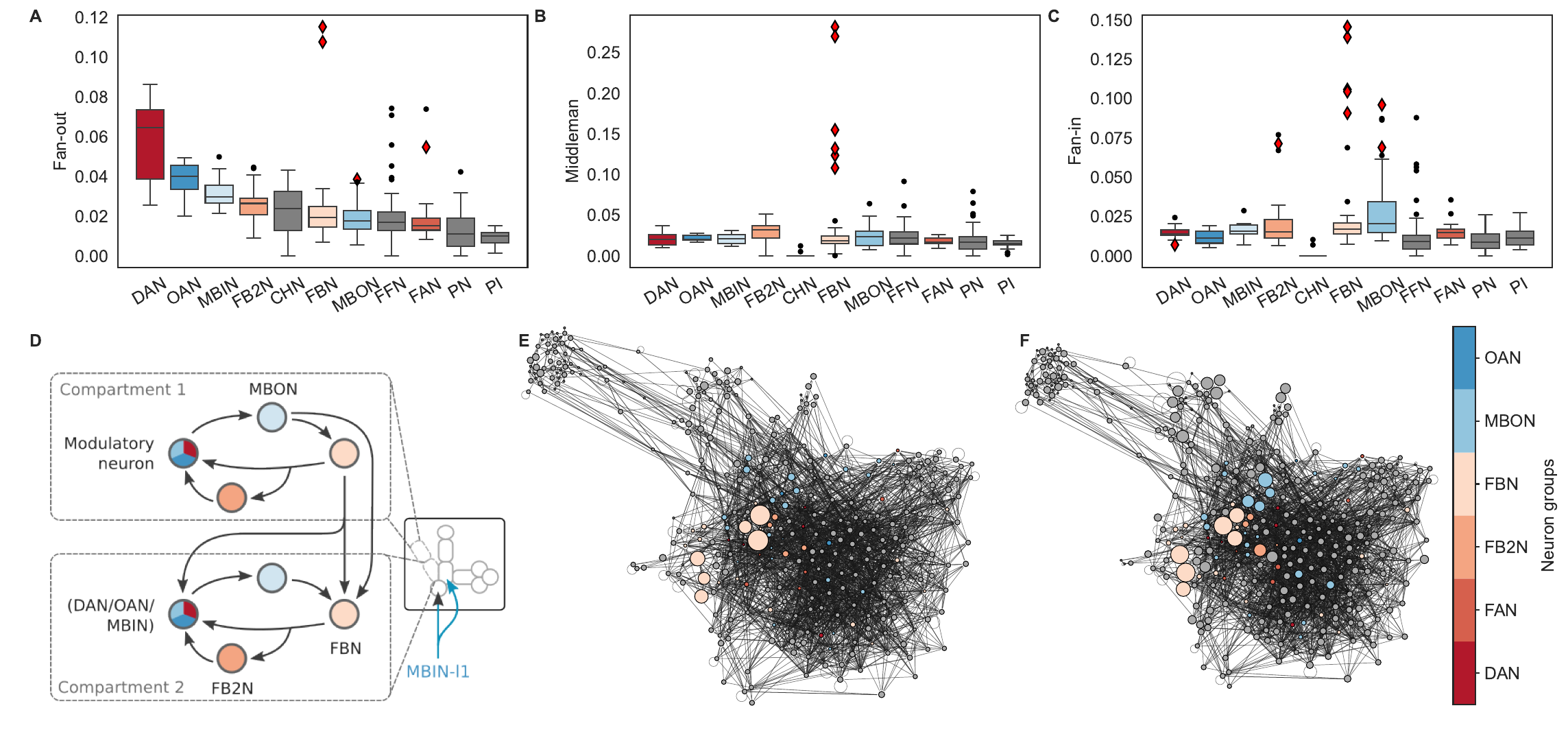}
	\caption{
		Directed and weighted clustering motifs identify both neuronal types and functional substructures in the drosophila larval mushroom body (MB). 
		\textbf{A}. The fan-out clustering distributions are clearly correlated with neuronal types, with the modulatory neurons (especially DAN and OAN) displaying significantly higher clustering with both continuous and Zhang--Horvath (not shown) methods.
		Neuron types are DAN, dopaminergic, OAN, octopaminergic, MBIN, modulatory input, FB2N, two-step feedback, CHN, chordotonal, FBN, feedback, MBON, modulatory output, FFN, feed-forward, FAN, feed-across, PN, projection, PI, inhibitory projection).
		Colored types indicate neurons that are involved in the recurrent circuitry of the network and are the same across all panels.
		\textbf{B--C}. No similar correlations are visible for the middleman and fan-in motifs where the distributions are dominated by outliers. 
		For middleman and fan-in motifs, neurons involved in the recurrent circuitry display clustering values that are slightly higher than those of the feed-forward and projection neurons.
		However the main visible feature is the presence of a few outliers with very high clustering values, which are mostly associated to neurons connected to the MBIN-l1 neuron (red diamonds).
		\textbf{D}. Structural organization of the MB into compartments with a proposed connectivity for the recurrent circuitry from \cite{eschbachRecurrentArchitectureAdaptive2020}, involving modulatory (DAN, OAN, MBIN), MBON, FBN/FAN, and FB2N units.
		The MBIN-l1 neuron is the only modulatory input neuron to innervate to distinct compartments.
		\textbf{E--F}. Force-Atlas representations of the network.
		For each plot, the node sizes represent the clustering values (from left to right: fan-out, middleman, fan-in).
		Coloured nodes in the two panels are neurons that are connected to MBIN-l1 and belong to one of the classes involved in the recurrent circuitry.
		These neurons display most of the largest clustering values in the network.
	}
	\label{fig:droso-modulatory}
\end{figure*}

We study different motifs on the connectome of the larval brain of the drosophila. We examine how various motifs (fan-in, fan-out and middleman) change for different neuron types.

We most prominently find that large fan-out values clearly distinguish the modulatory neurons (especially DAN and OAN) from the other neuronal classes (\supfigref{fig:droso-modulatory}A). 
As with the mouse connectome, we find that both fully weighted methods (continuous and Zhang–
Horvath) identify the correlation between the neuron types and the fan-out clustering fairly well. Nevertheless, in this case, we see that one of the hybrid methods (Onella) identifies the same pattern with the fully weighted methods, while Barrat's method failed to generate any pattern at all.

Furthermore, while the middleman and fan-in patterns do not show any strong correlation with any neuronal type, very high middleman and fan-in clustering values are found in neurons targeting the MBIN-l1, the only modulatory input neuron that innervates two distinct compartments. This reveals a strongly coupled structure between multiple modulatory neurons that appears to be more complex than the standard structure proposed in the original study.
These neurons include most of the highest clustering nodes, with 7 of the top 10 fan-in and middleman values.

\end{document}